\documentclass[]{spie}  

\usepackage{amsmath,amsfonts,amssymb}
\usepackage{graphicx}
\usepackage[colorlinks=true, allcolors=blue]{hyperref}
\usepackage{wrapfig}
\usepackage{soul}

\title{Effective Area calibration of the Nuclear Spectroscopic Telescope ARray (NuSTAR)} 

\author[a,b]{Kristin K. Madsen}
\author[c]{Karl Forster}
\author[c]{Brian Grefenstette}
\author[c]{Fiona A. Harrison}
\author[c]{Hiromasa Miyasaka}

\affil[a] {CRESST and X-ray Astrophysics Laboratory, NASA Goddard Space Flight Center, Greenbelt, MD 20771, USA}
\affil[b]{Department of Physics and Center for Space Science and Technology, University of Maryland, Baltimore County, Baltimore, MD
21250, USA}
\affil[c]{California Institute of Technology, 1200 E. California Blvd, Pasadena, USA}

\authorinfo{Further author information: (Send correspondence to K.K.M.)\\K.K.M.: E-mail: kmadsen@umbc.edu}

\newcommand{\nustar}{\emph{NuSTAR}} 
 
\newcommand{\as}{$''$}
\newcommand{\am}{$'$}

\newcommand{\nh}{$N_{\rm{H}}$}

\newcommand{\detabs}{\texttt{DETABS}}

\begin{document} 
\maketitle

\begin{abstract}

We present here the updated calibration of \textit{The Nuclear Spectroscopic Telescope ARray} (\nustar), which was performed using data on the Crab accumulated over the last 9 years in orbit. The basis for this new calibration contains over 250~ks of focused Crab (imaged through the optics) and over 500~ks of stray-light Crab (not imaged through optics). We measured an epoch averaged Crab spectrum of the stray-light Crab data and define a canonical Crab spectrum of $\Gamma = 2.103 \pm 0.001$ and N$ = 9.69\pm 0.02$~keV$^{-1}$ cm$^{-2}$ s$^{-1}$ at 1 keV, which we use as our calibration standard. The new calibration, released in the CALDB update 20211020, provides significant updates to: 1) the detector absorption component, 2) the detector response function, and 3) the effective area vignetting function. The calibration improves agreement between FPMA and FPMB across detectors with a standard deviation of 1.7\% for repeat observations between off-axis angles of 1\am--4\am, and the measured flux has increased by 5-15\%, with 5\% below 1\am\ off-axis angle, 10\% between 1-2\am, and 15\% above 4\am.  

\end{abstract}


\section{Introduction}

The \textit{Nuclear Spectroscopic Telescope Array} (\nustar) was launched in June 2012 \cite{Harrison2013} and  carries two co-aligned conical grazing incidence Wolter-I approximation \cite{Petre1985} optics modules (OMA, and OMB) that focus onto two identical focal plane modules (FPMA and FPMB). Each of these are composed of four solid state CdZnTe pixel detector arrays (enumerated Det0 through Det3). There are 133 shells in each optic, and the outer 43 shells are coated with a W/Si multilayer while the inner 90 shells are coated with Pt/C, limiting the highest efficient reflective X-ray energies to the Pt 78.4~keV K-edge \cite{Madsen2009}. 

The instrument has been in orbit for 9 years, and the original effective area calibration, reported in Madsen et al (2015) \cite{Madsen2015a}, was performed during the first year and a half of operation. Based on the results from Toor \& Seward (1974)\cite{Toor1974} and the investigation by Kirsh (2005)\cite{Kirsch2005}, we calibrated against a canonical Crab standard of $\Gamma \equiv 2.1$ and $N \equiv 8.5$~keV$^{-1}$ cm$^{-2}$ s$^{-1}$ at 1 keV. Then in 2016, there was an update of new detector gain corrections and the re-calibration of the detector absorption parameters, reported in Madsen et al. (2017) \cite{Madsen2017b}, which necessitated a readjustment of the vignetting function. Both of these effective area and vignetting calibrations were generated the same way, which we name the off-axis angle weighting method, or `oaa-weighting' for short and summarize below. In August 2020, and updated in July 2021, we provided a correction of the multilayer insulation (MLI) absorption covering OMA because evidence indicated that the MLI had been ripped and now exposes part of the optic aperture of OMA directly to space\cite{Madsen2020}. 

The original calibration and its 2016 update has generally performed well, but over the years features have emerged when observing high signal to noise sources in addition to those mentioned above:
\begin{itemize}
    \item differences between FPMA and FPMB spectra at low energy due to a mismatch between the detector absorption and the vignetting function,
    \item large differences in spectra from different detectors on the same FPM, related to the same mismatch between detector absorption and the vignetting function,
    \item low energy tails related to the detector response,
    \item absorption or line-like features in the spectrum related to the vignetting function.
\end{itemize}



Because the various components of the instrument response and their effects on the spectrum are interleaved, the re-calibration had to occur in a specific order. The calibration hinges upon the precise measurement of the stray-light Crab spectrum, and we first had to re-measure the correct detector absorption coefficient and address the low-energy detector response below 5~keV. We did this with the stray-light Crab data and the details are presented in Section \ref{detabsrmf}. Once those had been updated, we could proceed to the calibration of the telescope effective area and vignetting function using the focused Crab data, detailed in Section \ref{vignetting}.

In the following sections, we will refer to the collection of four detector as a module, and when we discuss individual detectors we will in short form designate them in the format Det0A, which would be Det0 on FPMA.

Readers interested only in the short short version with minimal details should read Sections: \ref{caldb}, \ref{crabspectrum}, \ref{avgcrab}, \ref{rmfcal}, \ref{focuscrabfit}, and \ref{result}.

\section{\texttt{NUSTARDAS} and the \texttt{CALDB}}\label{caldb}
A detailed summary of \texttt{NUSTARDAS} and \texttt{CALDB} revisions can be found on the \nustar\ Science Opeartions Center (SOC) pages\footnote{https://nustarsoc.caltech.edu/NuSTAR\_Public/NuSTAROperationSite/software\_calibration.php}, but we provide here a relevant summary for the analysis to be discussed. 

Until now, the RMF has not changed since its initial release in 2013-08-14, and we refer to this RMF as 2010v002, which reflects its version number in the \texttt{CALDB}. As part of the preparation to update the RMF, an intermediate RMF was created, which is described in Section \ref{rmfcal}, and we refer to this RMF as v3.0. This was an internal version to the SOC and was not released to the \texttt{CALDB}. The final corrected RMF product that is released to the \texttt{CALDB} we refer to as v3.1. This appears in the 20211020 \texttt{CALDB} as 2010v003.

The ARF, and in particular the vignetting function, has undergone a couple of iterations since launch. The version upon which the corrections described in this paper are built originates from 2016-06-06. We refer to this vignetting function as v007, based on its revision number in the \texttt{CALDB}. The final corrected vignetting file to be released to the 20211020 \texttt{CALDB} we refer to as v008.

In 2016-06-06, we also released \detabs\ v003, and the new updated version from Table \ref{nuabs} released in the 20211020 \texttt{CALDB} we refer to as \detabs\ v004.

For all data reductions that use \texttt{NUSTARDAS} we use version v2.1.1.

\section{\nustar\ calibration methodology}\label{methodology}

Preferably instruments should be understood by their ground calibration and only require additional in-orbit calibration to verify the ground results. Reality, however, has shown that instruments rarely perform exactly as they did on the ground. Sometimes this is due the challenges involved in translating calibration taken at facilities to infinitely distant point sources on the sky, or it can be due to a process that happened after the ground calibration occurred. In some cases, corrections can be made based on an understood physical process, such as a time dependent contamination layer or detector gain shifts. But sometimes the underlying physical process, or geometry, may not be known or too complex to accurately model. 

For \nustar, the fundamental challenge in calibrating the optics resides in the complexities of the multilayer response. There are 10 different multilayer recipes\cite{Madsen2009}, approximately evenly distributed across $\sim$3000 individual segments of glass, and although the theoretical responses of the recipes are known, and we know where each segment resides, during production variations occurred in the recipes. Since we did not measure the responses of all individual pieces, which are highly dependent on grazing incidence angle, we do not know the true response of each piece. Deviations from the theoretical response of the full optic were therefore expected and directly related to the differences observed in the multilayers. Unfortunately, because the parameter space is so large (3000 individual mirrors times their response as a function of off-axis angle) it is not feasible to solve for the individual multilayer responses, and predicting the response from a raytrace simulation consequently not accurate enough.

It was therefore necessary to calibrate the effective area of the \nustar\ optics against a standard candle. We used the Crab, which  is a center filled pulsar wind nebula (PWN) powered by a pulsar with a double peaked profile of period P $\sim$ 33\,ms. It has served as the primary celestial calibration source for many hard X-ray instruments because of its brightness, relative stability, and simple power-law spectrum over the band from 1--100\,keV \cite{Kirsch2005}. The spatially phase-averaged integrated spectrum of the Crab nebula+pulsar in the 1--100\,keV X-ray band has been well-described by a power-law with photon index of $\Gamma\sim 2.1$ \cite{Kirsch2005}, while above 100\,keV the hard X-ray instruments (\textit{INTEGRAL}/SPI/ISGRI, \textit{CGRO}) measure a softer index of $\Gamma \sim $2.20--2.25. The exact location of the turnover is not well determined. It appears very gradual and estimated to occur somewhere between 50-120~keV \cite{Jourdain2009}. Weisskopf et al (2010)\cite{Weisskopf2010} did a study with \textit{RXTE}/PCA, \textit{XMM}, and \textit{ASCA}, looking for deviations from a power-law in the 0.2--50\,keV energy range, and concluded that within the precision of the available instrumentation there is no detectable bend in the phase-averaged integrated spectrum of the Crab. Measurements using \nustar\ stray-light data support these findings\cite{Madsen2017b}. The stability of the Crab has been tracked in great detail over the last decade and during this period the flux has been observed to change by 7\% across the 10--100\,keV bandpass \cite{Wilson2011}. The change, however, is slow, and the deviation in flux per year over the period it has been observed on the order of $\sim$3\%. The spectral index has also been observed to vary peak-to-peak by $\Delta \Gamma \sim 0.025$ in \textit{RXTE}/PCA \cite{Shaposhnikov2012}, but, as with the flux variation, it is slow. On average over the 16 years the Crab has been observed, it has remained steadily at  $\Gamma \sim 2.1$. 

The task of the \nustar\ calibration is thus to compare the measured spectrum to the assumed Crab model. We are unable to account for the intrinsic variations within the source over time and must treat them as systematic. However, since the systematic errors of the instrument from repeated measurement are of a similar magnitude\cite{Madsen2015a}, the intrinsic variability of the Crab is acceptable for our purpose.

\subsection{The Vignetting Correction Function}\label{corrfunc}

Mathematically, we model the detected counts in a given instrumental pulse height bin, $C(PI,\theta)$, according to the
equation,
\begin{eqnarray}
C(PI,\theta) &=& \int{\frac{dN(E)}{dE} R(PI,E,\theta) dE}\,.
\end{eqnarray}
Here $dN(E)/dE$ is the model differential photon spectrum of the observed target as a function of incident photon energy, $E$, and $R(PI,E,\theta)$ is the instrument response that captures the incident photon in a given pulse height bin, $PI$, at an off-axis angle, $\theta$. In practice, this integral is approximated as a finite sum by sampling $R(PI,E,\theta)$ on a grid.  The off-axis angle, $\theta$, is the angle of incoming X-rays with respect to the optical axis of the telescope.  As the optical axis moves with respect to the detector position during an observation, the modeled response is sampled on a finite time grid and then summed for a given exposure.

As is typical for X-ray astrophysical missions, the response matrix is divided into two components,
\begin{eqnarray} 
R(PI,E,\theta) &=& {\rm RMF}(PI,E) \circledast A(E,\theta).
\end{eqnarray}
RMF$(PI,E)$ is known as the redistribution matrix, which contains detector quantum efficiency and resolution effects, and it is unitless (a fraction between 0 and 1, where a value of 1 indicates 100\% quantum efficiency). The RMF will be discussed in greater detail in Section \ref{rmfcal}. The quantity $A(E,\theta)$ is the effective area, also known as the ancillary response function (ARF), which captures the effective area of the optics as well as several other attenuation factors unique to \textit{NuSTAR}:
\begin{eqnarray} 
A(E,\theta) &=& A_o(E) \times V(E,\theta) \times {\rm DETABS}(E) \times {\rm GR}(E,\theta)\times  {\rm AS}(E,\theta) \times Corr(E,\theta)\,.
\end{eqnarray}
Here $A_0(E)$ is the modeled on-axis effective area of the mirror segments estimated from theoretical ray-tracing simulations, and $V(E,\theta)$ the geometric vignetting function, also based on ray-tracing simulations. The quantities \detabs, GR, and AS are the detector dead layer absorption (discussed in Section \ref{detabs}), ghost ray correction, and aperture stop correction. We do not concern ourselves with the GR and AP, and details on these components can be found in Madsen et al (2015)\cite{Madsen2015a} and Madsen et al (2017)\cite{Madsen2017a}. For the ARF, $Corr(E,\theta)$ is the empirically derived correction factor we want to obtain.

In practice, discovering the correction function $Corr(E,\theta)$ involves taking numerous observations of the Crab at different off-axis angles. In this way a grid of countrates at $E$ and $\theta$ are obtained and can be used to interpolate the response at any desired point, where, obviously, the finer the grid the more accurate the function.

\begin{figure}
\begin{center}
\includegraphics[width=0.40\textwidth]{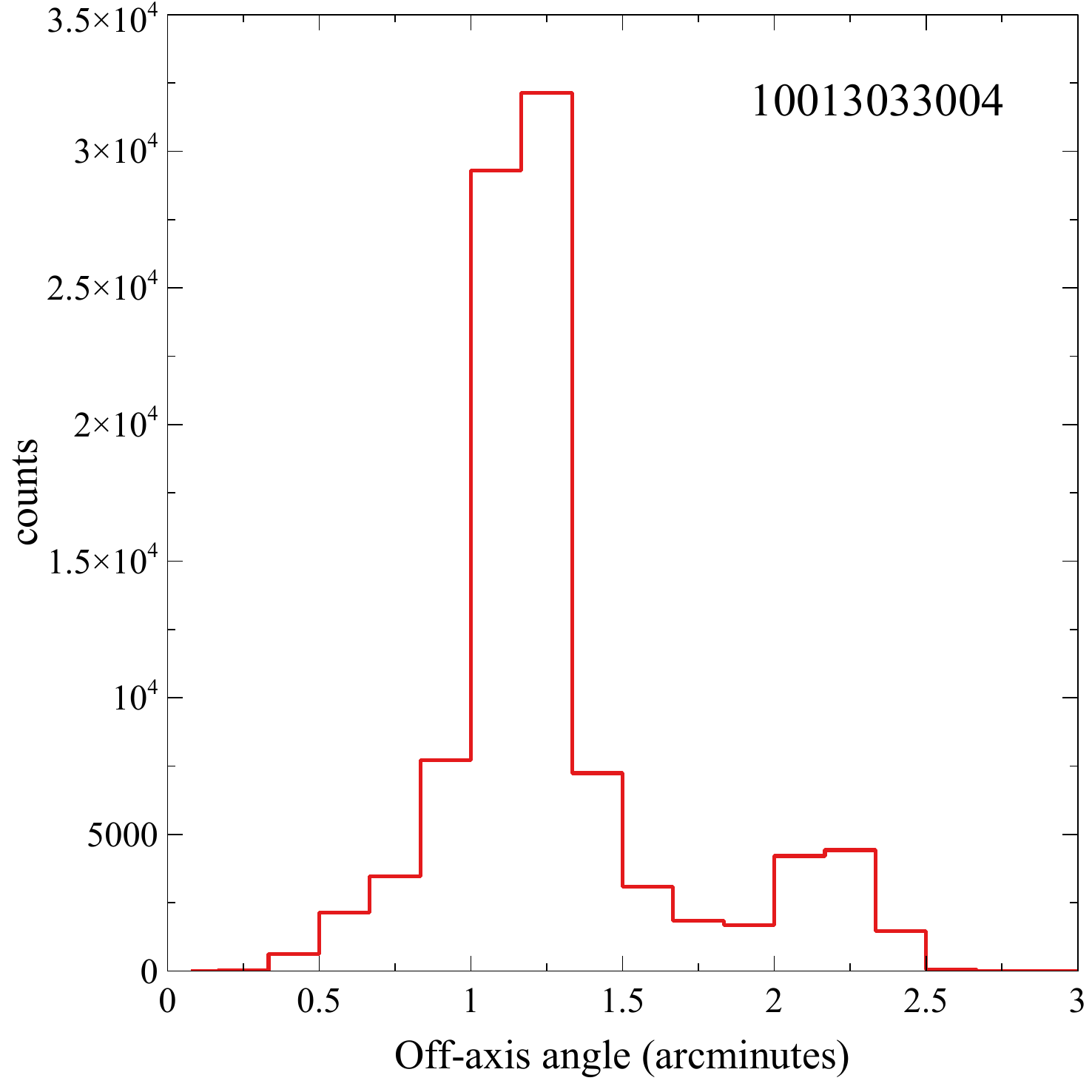}
\includegraphics[width=0.49\textwidth]{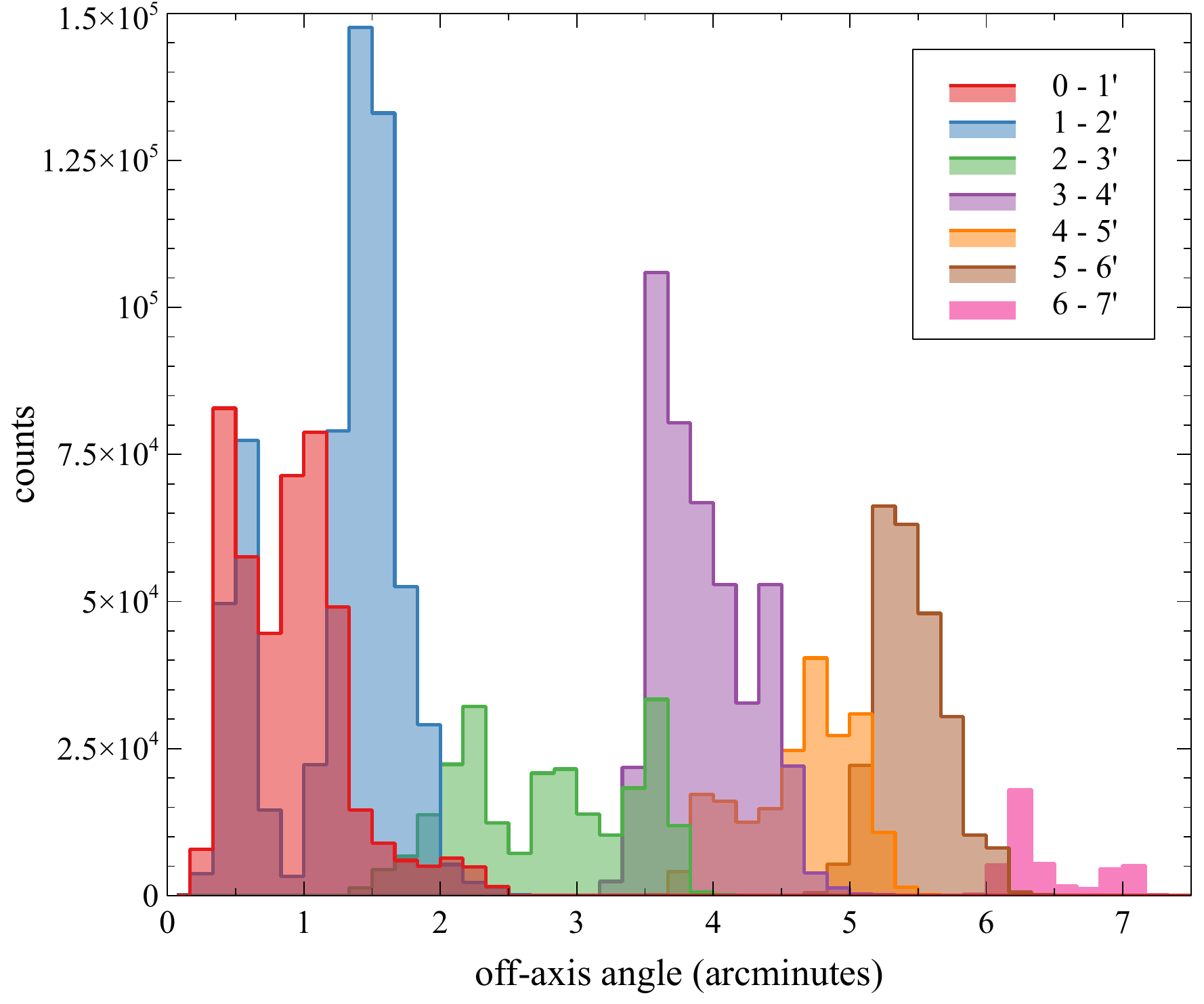}
\end{center}
\caption{Left: Histogram of the count distribution between 3-78~keV as a function of off-axis angle for a single observation in 10\as\ bins. Right: Histogram of the 'oaa-weighting' of the original binning using the $\sim 35$ observations available at the time \cite{Madsen2015a}.}
\label{oldoaahisto}
\end{figure}

With \nustar, however, obtaining such a data set is problematic. The observatory has an optics bench and a focal plane bench separated by 10.14 meters, which move relative to each other due to motions of the mast that connects them and causes the optical axis to travel across the FPMs by up to several arcminutes. This means that each observation samples a range of off-axis angles that is unique for that observation, as shown by Figure \ref{oldoaahisto} (left) for observation 10013033004. In this particular observation, the off-axis angle distribution covers more than 2\am~and, except for a narrow range, there is not sufficient signal to measure the spectrum accurately as a function of off-axis angle. To overcome this problem, we add together many such observations to increase statistics in each bin. In our original oaa-weighted method we did this by calculating the mean off-axis angle for each observation and assigning the observations to one arcminute bins: 0-1\am, 1-2\am, ..., 6-7\am. For each bin the assigned observations were then combined, and a new off-axis angle for the combined effective area was calculated by weighting the individual off-axis angle distributions. 

\begin{figure}
\begin{center}
\includegraphics[width=0.35\textwidth]{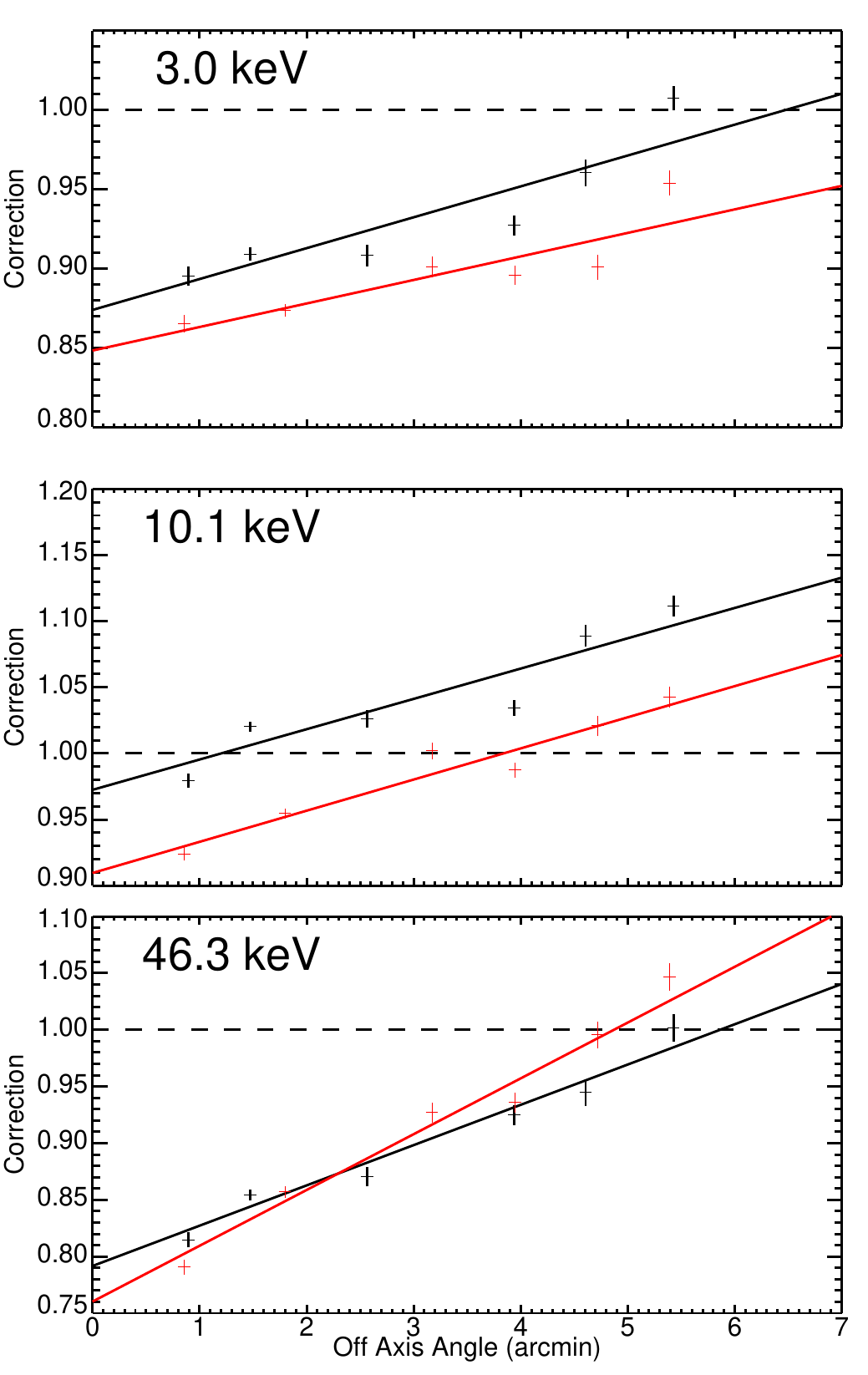}
\includegraphics[width=0.55\textwidth]{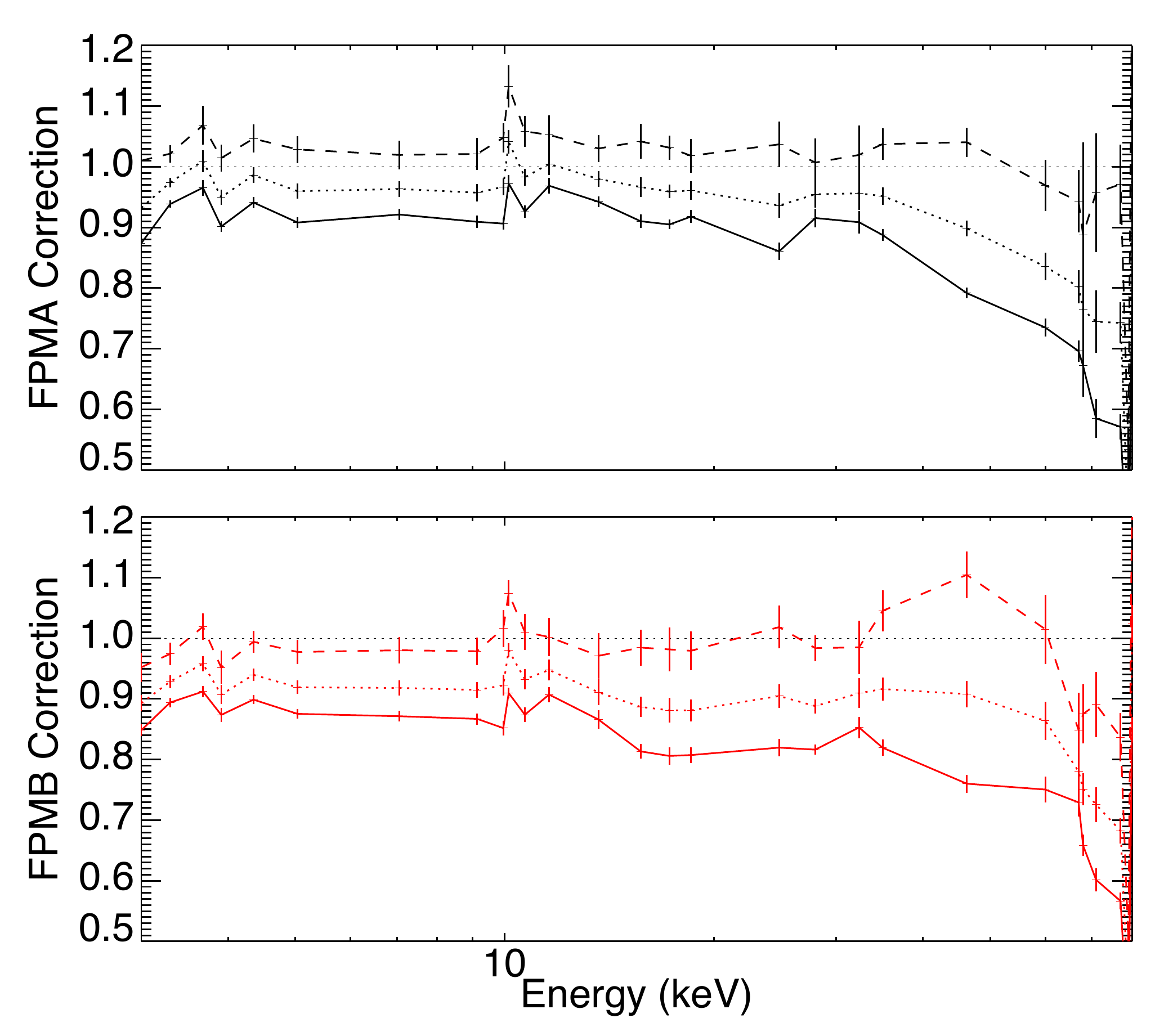}
\end{center}
\caption{Correction functions, $C(\theta,E)$, of the previous \nustar\ calibration\cite{Madsen2015a}. Left: The linear (in $\theta$) function for selected energies. Right: $C(\theta,E)$ for offaxis angles 0\am\ (solid), 3\am\ (dotted), and 7\am\ (dashed).}
\label{oldcorrfunc}
\end{figure}

The issue with this method is obviously that the combined distributions are overlapping as shown in Figure \ref{oldoaahisto} (right), and that the spectra are sampling a range of off-axis angles that go beyond the assigned bin. Also, because the bins were required to be large enough to accumulate enough statistics, our knowledge of what happened in between is poor. For this old dataset, Figure \ref{oldcorrfunc} (left) shows the effective area correction factor, $C(E,\theta)$ for energies 3.0, 10.1, and 46.3~keV. Although to first approximation the correlation appears linear as a function of off-axis angle, there is residual substructure clearly visible.

\subsubsection{New Method}
\label{newmethod}

To address the issues of the previous calibration, we devised a new method. The number of Crab observations has grown from the initial 39 to 71. With more than double the increase in exposure time this allows for a higher fidelity analysis than was previously available. The most important change is that we calculate the off-axis angle of each individual photon. Due to the extent of the Crab and the PSF, we do not measure the actual photon's position, but the off-axis angle of the source center at the time of the photon's arrival. In this manner, we can assign each photon to a specific off-axis angle bin without having overlapping distributions. 

\begin{figure}
\begin{center}
\includegraphics[width=0.48\textwidth]{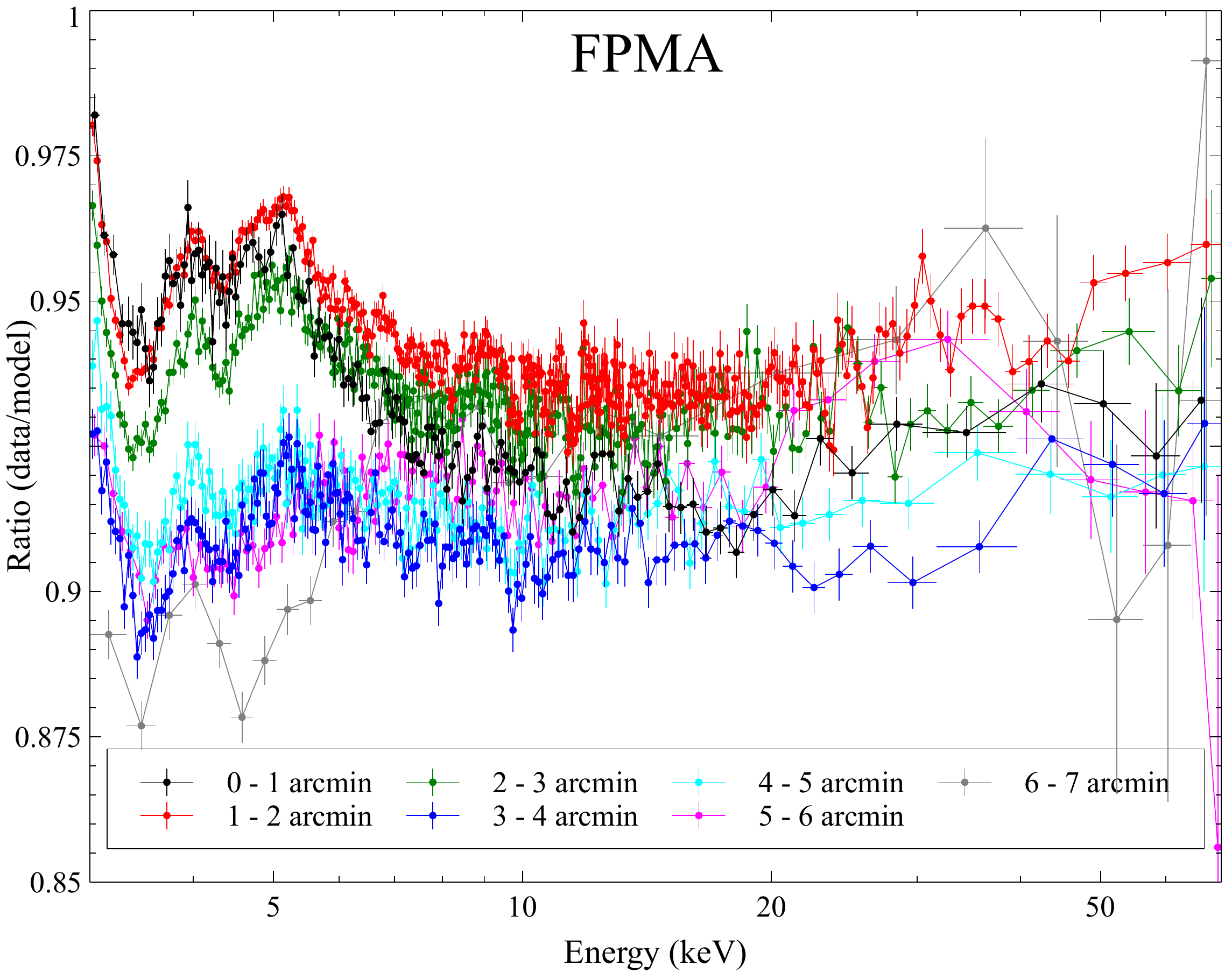}
\includegraphics[width=0.48\textwidth]{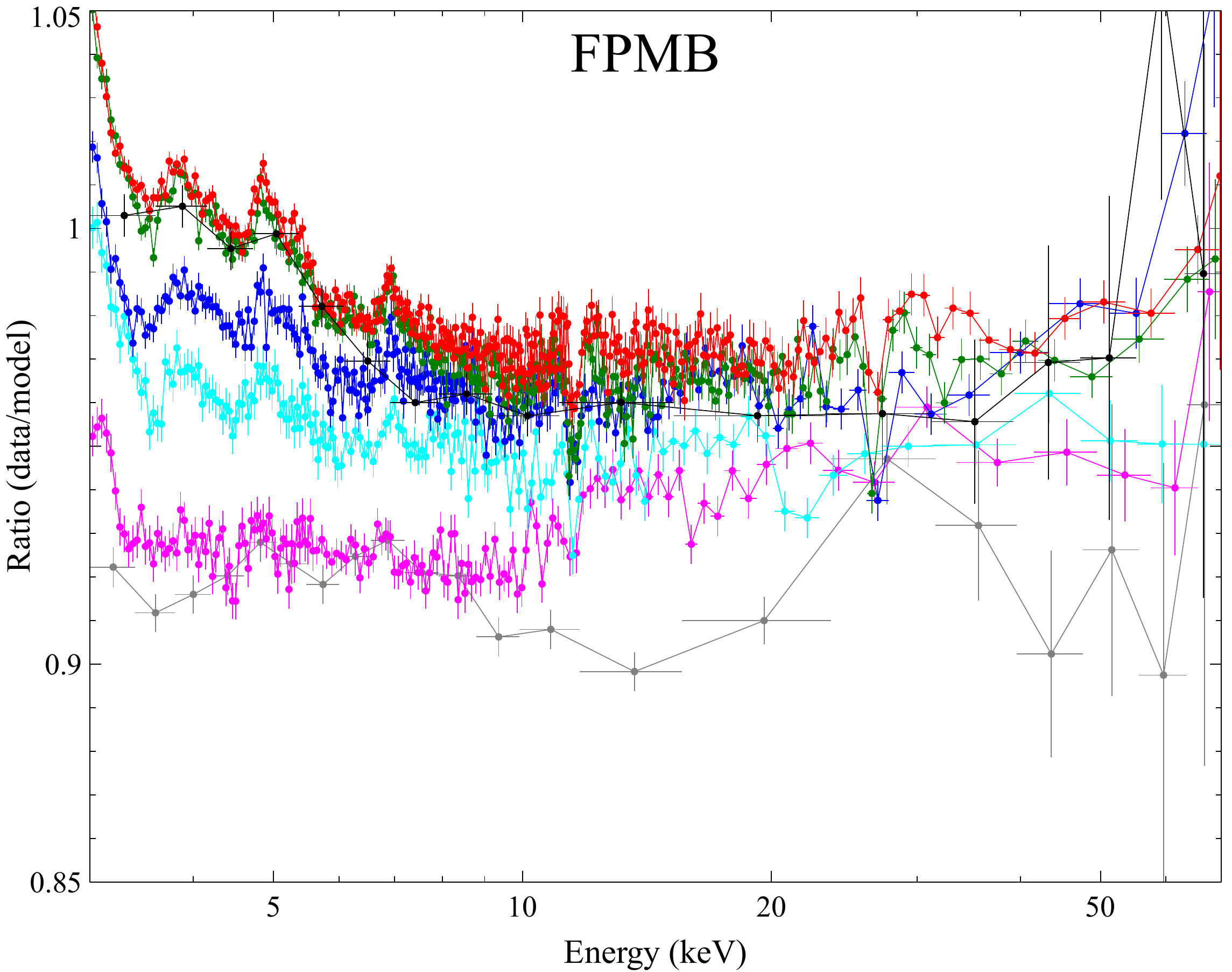}
\end{center}
\caption{Ratio of the data reduced with ARF v007, but with new detector absorption parameters (see Table \ref{detabs}) and RMF v3.1, to the Crab model in 1 arcminute bins for FPMA and FPMB. This shows that the flux increase will be largest for larger off-axis angles and low energies. The abrupt change in FPMA above 3\am\ is because at these off-axis distances the source is typically on another detector that has had  poorly constrained detector absorption parameters.}
\label{ratio1am}
\end{figure}

\begin{figure}
\begin{center}
\includegraphics[width=0.48\textwidth]{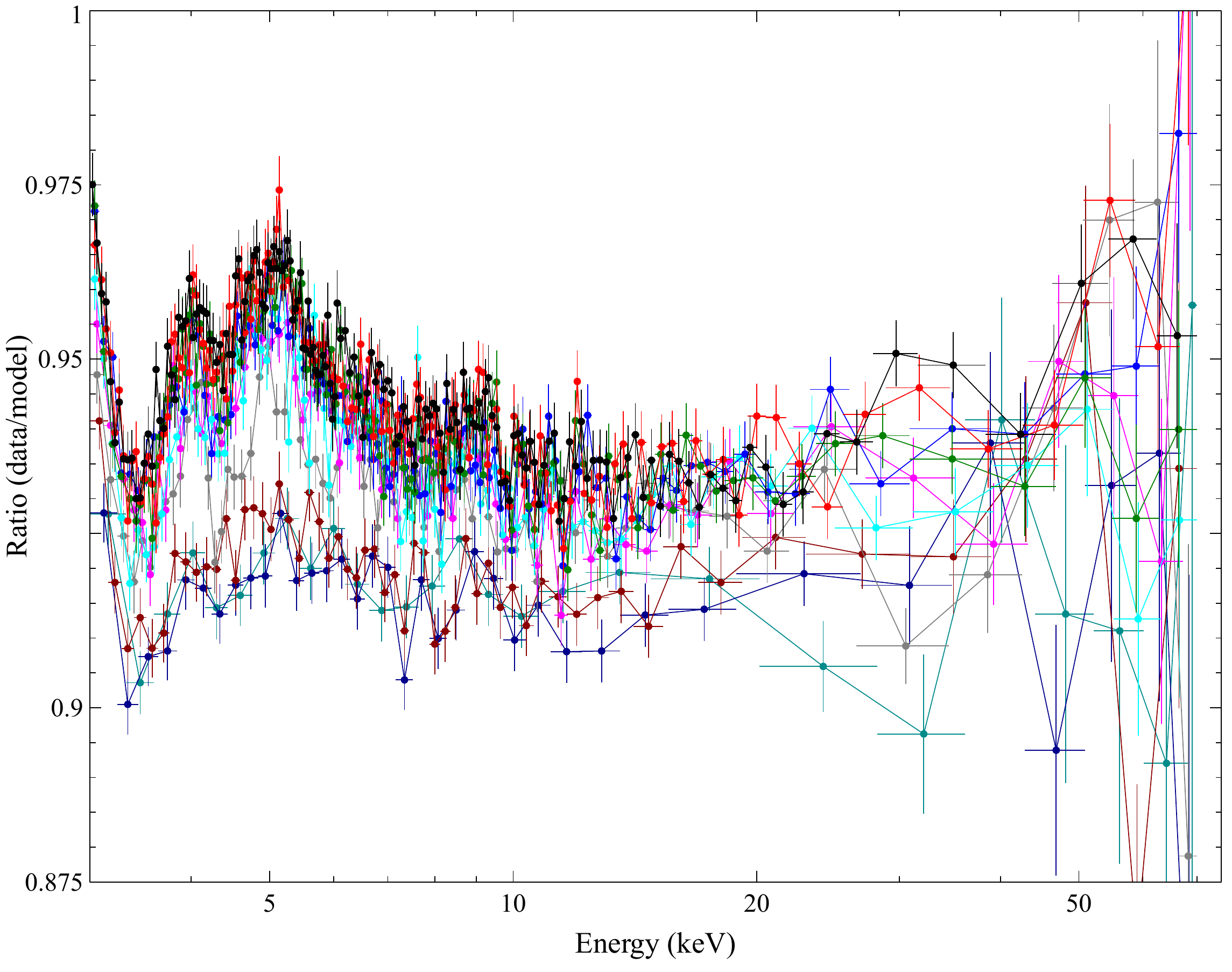}
\includegraphics[width=0.48\textwidth]{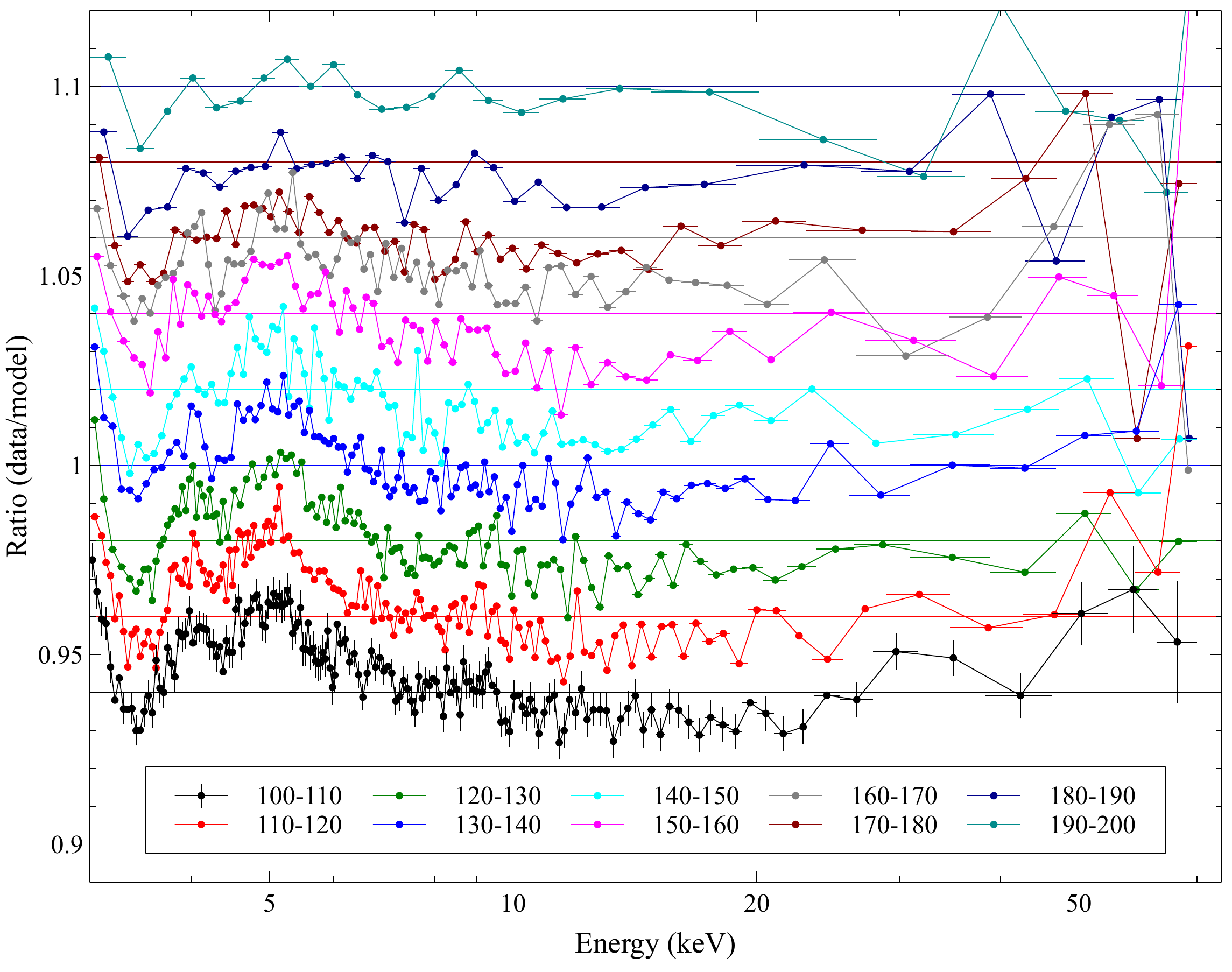}
\end{center}
\caption{FPMA ratio of the data to the Crab model for off-axis bins 100-200\as. Left: The curves maintained at their absolute value. Right: the curves have been offset by 0.02 and error bars removed for clarity. Systematic variations from bin to bin are evident.}
\label{ratio10as}
\end{figure}

Since the angular resolution of the response files is 10\as, we use 10\as\ as the smallest subdivision. We also bin at 60\as, which becomes useful for higher energies and large off-axis angles. We combine all observations from the same off-axis angle bin together in an epoch-mixed spectrum, which we can then fit against the reference Crab spectrum to obtain the correction for each 10\as\ bin instead of each arcminute. An example of this is shown in Figure \ref{ratio1am} for FPMA, where the ratio spectra are binned in 1\am\ and Figure \ref{ratio10as} the ratio spectra in 10\as\ bins between off-axis angles 100-200\as. Systematic, non-linear changes across the sub-arcminute bin can be observed. However, we note that the large deviations from the Crab spectrum shown here, particularly at low energies, are far more severe looking than would be obtained with the original responses. The reason for this is that the adjustments to the detector absorption parameters and the new RMF to be discussed in Section \ref{detabsrmf} must be included in the data reduction before deriving the corrections to the vignetting function itself. This in particular affects the low energies, and these ratio plots therefore indicate the level to which the detector related instrument terms were previously incorrectly included into the vignetting function. In short, much of what follows is primarily a bookkeeping exercise designed to correctly assign the response changes to the detectors or the vignetting.

\section{\nh\ of The Crab spectrum}\label{crabspectrum}
In the \nustar\ band the Crab spectrum is well represented by a powerlaw with galactic absorption expressed in \textsc{XSPEC} nomenclature as \texttt{tbabs $\times$ powerlaw}. For the \nh\ in \texttt{tbabs} we use Wilms \cite{Wilms2000} abundances and Verner\cite{Verner1996} cross-sections. The \nh\ of the Crab has over the years taken on a range of values as measured by different observatories, but is mostly constrained between $2\times 10^{21}~\mathrm{cm}^{-2}$ to 6$\times 10^{21}~\mathrm{cm}^{-2}$, with an average value of $\sim 4 \times 10^{21}~\mathrm{cm}^{-2}$ \cite{Kirsch2005}. In the original \nustar calibration in 2013 we measured the Crab column to be \nh=$2.2 \pm 2.0 \times 10^{21}\mathrm{cm}^{-2}$ \cite{Madsen2015a}. Since this component will be degenerate with some of the instrument components, we decided to maintain this value, and in everything that follows the \nh\ will be frozen at \nh=$2.2 \pm 2.0 \times 10^{21}~\mathrm{cm}^{-2}$. 

\begin{figure}
\begin{center}
\includegraphics[width=0.70\textwidth]{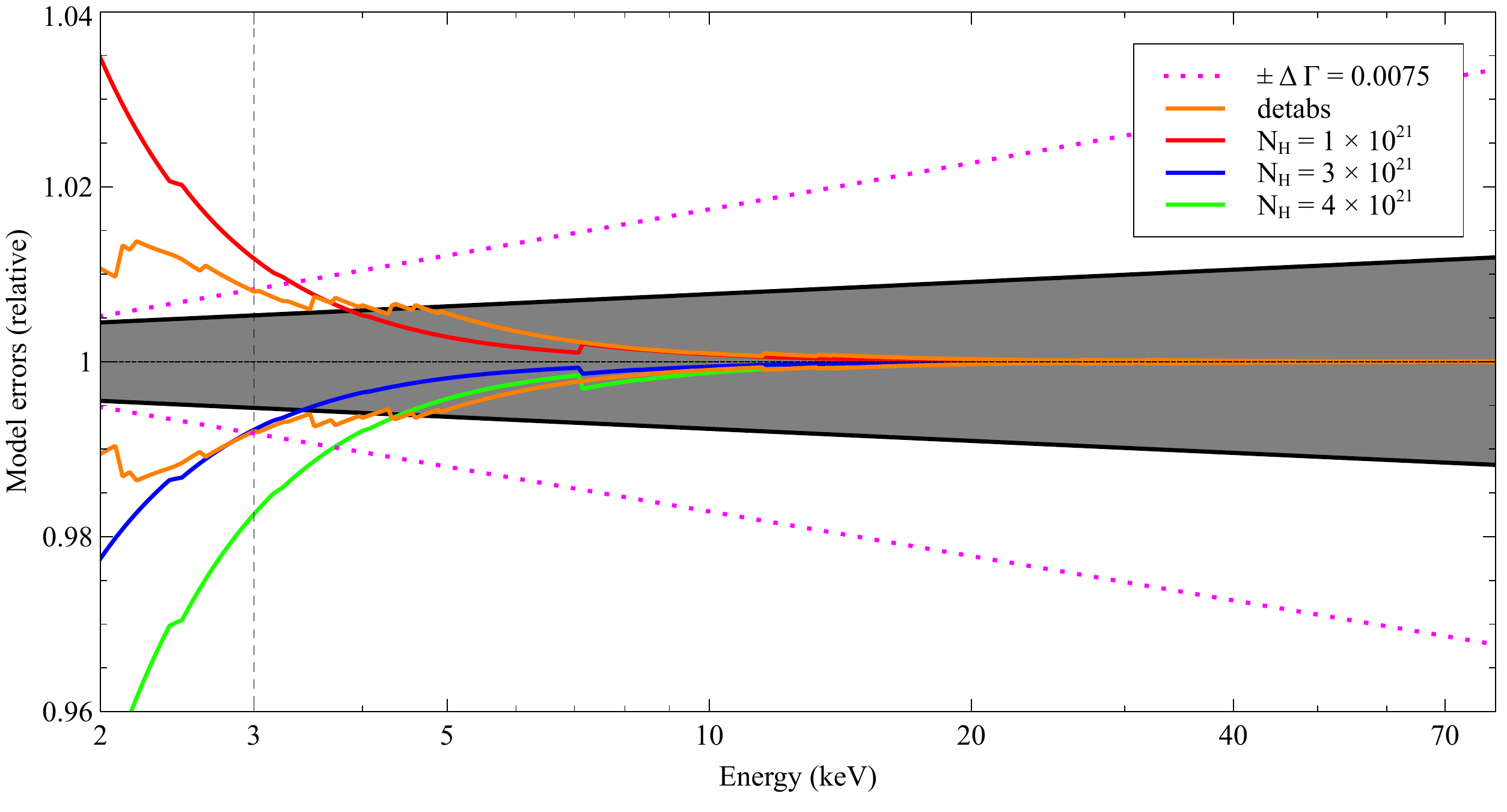}
\end{center}
\caption{Relative model components in the Crab fit. Shaded grey area is the fit obtained in Section \ref{avgcrab} bounded by the 1$\sigma$ errors. The \detabs\ curves are 1$\sigma$ deviations to the parameters for Det0A in Table \ref{nuabs}.}
\label{modelplot}
\end{figure}

Figure \ref{modelplot} shows a range of \nh\ and the impact it would have if it were different than presumed. At the \nustar\ lower calibration boundary of 3~keV, this could potentially account for a 1-2~\% error, which will be transferred into the instrument response. This translates into a systematic uncertainty on \nh\ that will be on the order of the measured spread in Crab \nh. The typical statistical errors in fitting \nh\ down to 3~keV in \nustar\ are of the order $\sim 1 \times 10^{21}\mathrm{cm}^{-2}$.

\section{Detector Absorption and RMF calibration}\label{detabsrmf}
\begin{figure}
\begin{center}
\includegraphics[width=0.70\textwidth]{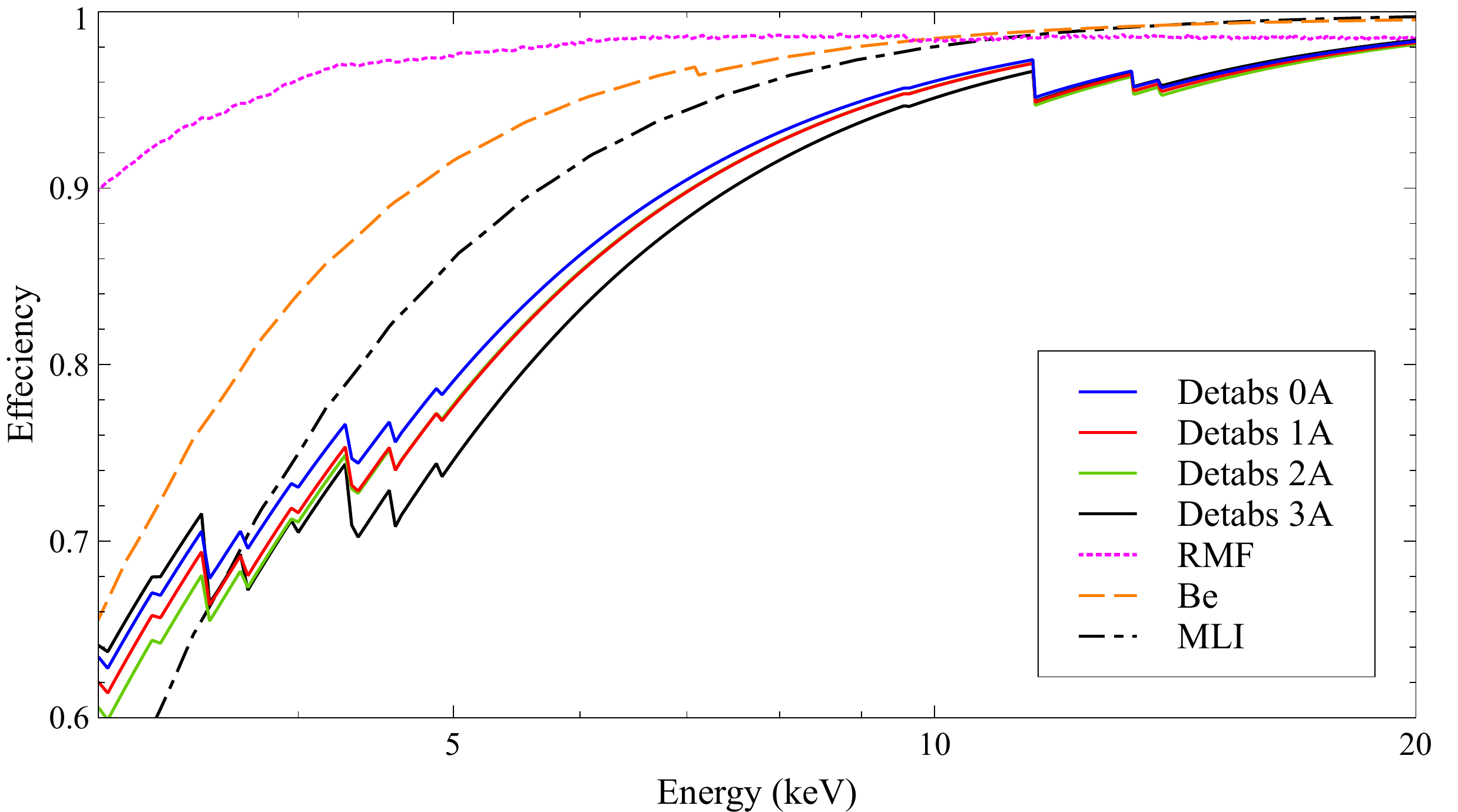}
\end{center}
\caption{Low-energy absorption components for FPMA. Values for FPMA \detabs\ given in Table \ref{nuabs}}
\label{effplot}
\end{figure}

Below $\sim$5~keV the instrument response is composed of several partially degenerate absorption effects as shown in Figure \ref{effplot}. Starting from the sky-side, the photon first encounters a layer of MLI before entering the optics and exiting again through another layer of MLI. The complications and challenges of calibrating the MLI are discussed in Madsen et al. (2020)\cite{Madsen2020}, but to summarize: the thickness of the MLI layer was absolutely calibrated on the ground pre-launch, and in-orbit re-calibration is only possible on relative differences, which we performed for FPMA after  discovering that one of the layers had ripped. When the photon reaches the focal plane bench, it enters the FPMs through a Be window, which like the MLI was calibrated  on the ground. We have not performed a re-calibration of this window and do not intend to as the only change to this calibration is if it gets punctured by a micrometeorite of which there is no evidence. At the detector surface, a Pt contact coating and an inactive CdZnTe dead layer cause absorption before the photon interacts in the active region of the detector. Unlike the MLI and Be window, no ground calibration of the dead layer was performed. The thickness of these layers, collectively referred to as detector absorption or \detabs, is measured in-orbit under the assumption that the MLI and Be thicknesses are constant and unchanging which is a good assumption in the absence of micrometeorite strikes. Finally, the detectors lose some quantum efficiency across the same energies due to electronic and charge transport effects.

Because all of these components have a similar effect on the spectrum, they are not easy to separate. During the previous calibration the detector absorption (\detabs) and the detector efficiency (\texttt{RMF}) became partially tangled with the vignetting function; the correction to the vignetting was used to collectively correct inaccuracies across all sub-components that, at the time, had insufficient data to be corrected individually. After nearly 10 years in orbit, we now have the capability to perform this task. In this section we use Crab stray-light (SL) observations\cite{Madsen2017a,Madsen2017b}, which are observations where the source flux bypasses the optics and falls directly on the detectors, to remove the degeneracy from the vignetting function. 

\nustar\ has a legacy program monitoring the Crab in the SL configuration. We use this data from several epochs to measure the thickness of the Pt and CdZnTe layers between 3--40~keV. Once we have those values we can the adjust the RMF between 2.2--5 keV. The valid lower energy bound of \nustar\ remains at 3~keV, but we calibrate down to 2.2~keV to ensure that the detector redistribution within the RMF does not cause problems at 3~keV. The Crab spectrum itself is fitted at the same time and gives us an absolute Crab flux reference. The time-averaged Crab spectrum over all epochs obtained from this data set is then used in the subsequent vignetting correction as the absolute reference.

\subsection{Observations and Data Reductions}
\begin{figure}
\begin{center}
\includegraphics[width=0.70\textwidth]{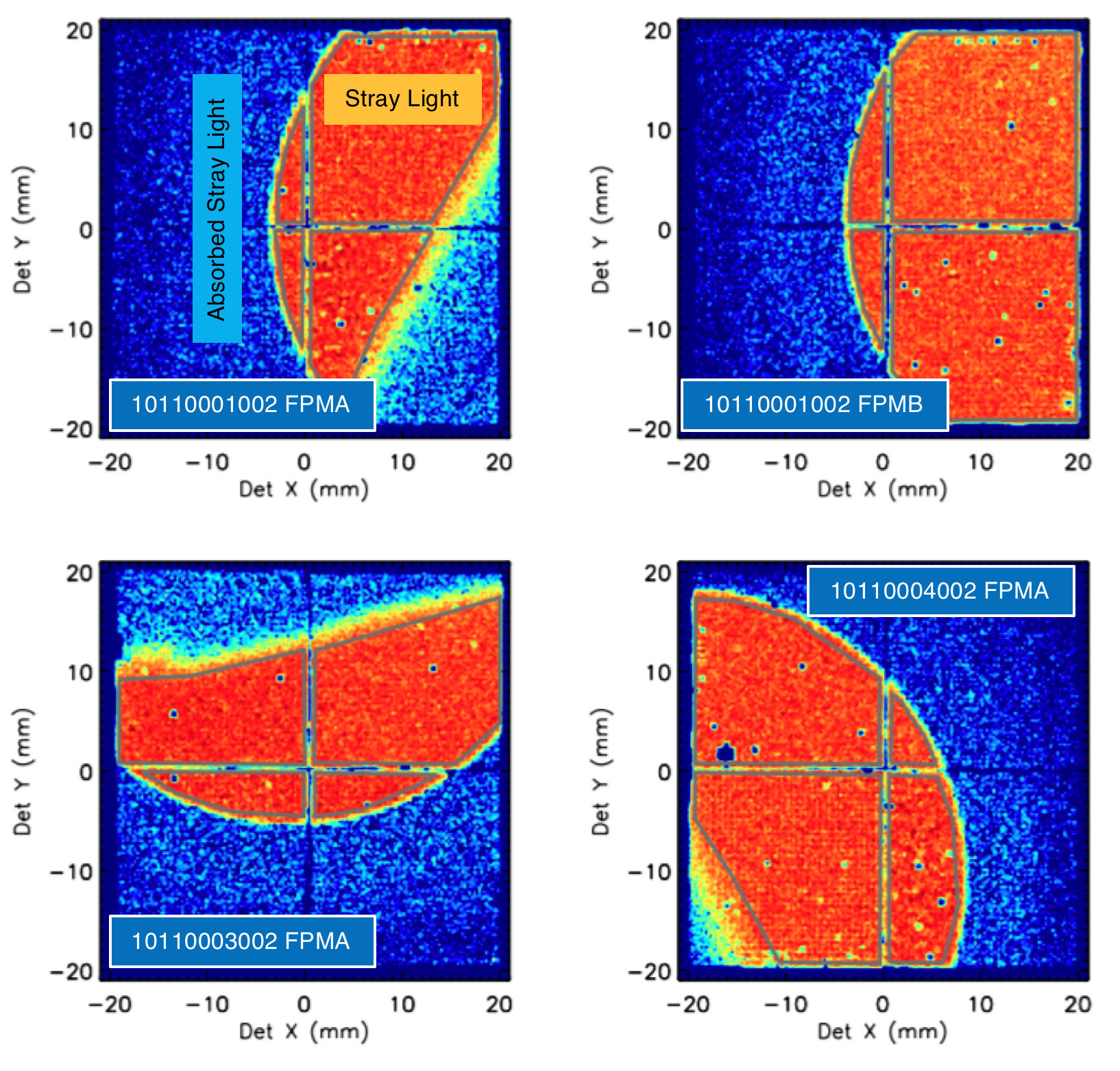}
\end{center}
\caption{Contour color plots of the detectors with logarithmic scaling. Green polygons show extraction regions, which are done for each detector individually. Due to the partially absorbed stray-light, backgrounds can not be taken from the same observation where there is SL.}
\label{SLexample}
\end{figure}

Stray-light observations are performed by placing the source at $\sim1$~deg off-axis, which maximises the amount of stray-light so that it roughly covers half the FoV. If the source is moved closer it gets blocked by the optical bench, while if it is moved further away less stray-light falls on the detector. Figure \ref{SLexample} shows examples of stray-light from the Crab, some of which are starting to be blocked by the optical bench. Due to the geometry of the observatory, at most two detectors can be covered at a time. The circular shape comes from the circular cutout in the aperture stops. In addition to the regular stray-light there is a partially absorbed stray-light component as well (see Figure \ref{SLexample}, top left). This is source flux passing through the aperture stops. Since this is always present in the area adjacent to the main SL it precludes taking a background from the same observation. For this reason the legacy Crab stray-light observations are accompanied by ``blank sky" background observations taken at $\sim10$~deg from the Crab. Table \ref{SLobslog} lists all the stray-light observations used here with their background observations.  

The stray-light data are reduced with \texttt{nupipeline} using default settings and the background filters set to the appropriate parameters as evaluated from the background filtering diagnostic pages\footnote{https://nustarsoc.caltech.edu/NuSTAR\_Public/NuSTAROperationSite/SAA\_Filtering/SAA\_Filter.php}. We do not use \texttt{nuproducts} for spectral extraction, but created custom tools to extract the data in detector coordinates \cite{Madsen2017a}. These tools have since been ported into python for general use and are described in more detail in Grefenstette et al. (2021)\cite{Grefenstette2021}. Here, however, we use the original tools, which allow for greater flexibility and adjustment. In short, the procedure creates an ARF, which only contains the Be absorption from the FPM entrance window, that is scaled by the geometric area illuminated by source on the detector and adjusted by the dead pixel area. As shown in Figure \ref{SLexample}, the SL spectra are extracted detector-by-detector since the \detabs\ parameters are distinct to the individual detectors and these are the values we want to recover. Since the spectra are only taken from individual detectors we can directly use the intermediate RMF v3.0 described in Section \ref{rmfcal}. 

\begin{table}
\centering
\caption{Stray Light Observing Log}\label{SLobslog}
\begin{tabular}{l|cccll}
\hline
Epoch & Obsid & Bkg Obsid & exposure time (sec) & Module list$^a$ & Detector list$^b$ \\
\hline
\hline
    Oct 2015 & 10110001002  &  10110002002 & 18961  &  A,A,A,A  &     0,1,2,3 \\
    Oct 2015 & 10110001002  &  10110004002 & 18961  &  B,B,B,B  &     0,1,2,3 \\
    Oct 2015 & 10110003002  &  10110004002 & 19299  &  B,B  &                 0,1 \\
    Apr 2016 & 10110004002  &  10110005001 & 20888  &  A,A,A  &               1,2,3 \\
    Apr 2016 & 10110005001  &  10110004002 & 21776  &  B,B  &                 0,1 \\
    Feb 2017 & 10210001002  &  10210002002 & 20173  &             A,A,B,B  &             0,3,0,3 \\
    Apr 1017 & 10210001003  &  10210002003 & 18158  &             A,A,B,B  &             0,3,0,3 \\
    Jul 2017 & 10311001002  &  10311002004 & 17305  &             A,A,B,B  &             0,3,0,3 \\
    Sep 2017 & 10311001004  &  10311002004 & 17911  &             A,A,B,B  &             0,3,0,3 \\
    Oct 2017 & 10311001006  &  10311002006 & 20442  &             A,A,B,B  &             0,3,0,3 \\
    Nov 2017 & 10311001008  &  10311002008 & 19467  &             A,A,B,B  &             0,3,0,3 \\
    Mar 2018 & 10402002002  &  10402006002 & 19133  &             A,A,B,B  &             0,3,0,3 \\
    Mar 2018 & 10402002004  &  10402006002 & 20848  &             A,A,B,B  &             0,3,0,3 \\
    Mar 2018 & 10402002006  &  10402006002 & 20085  &             A,A,B,B  &             0,3,0,3 \\
    May 2018 & 10402008002  &  10502004002 & 13877  &             A,A,B,B  &             0,3,0,3 \\
    Sep 2018 & 10402002002  & 10402003002 &  19280  &             A,A,B,B  &             0,3,0,3 \\
    Sep 2018 & 10402002004  & 10402003002 &  20848  &             A,A,B,B  &             0,3,0,3 \\
    Sep 2018 & 10402002006  & 10402003002 & 20085  &              A,A,B,B  &             0,3,0,3 \\
    Feb 2019 & 10502002002  &  10502004002 & 28704  &             A,A,B,B  &             0,3,0,3 \\
    Mar 2019 & 10502002004  &  10502004002 & 28705  &             A,A,B,B  &             0,3,0,3 \\
    May 2019a & 10502006002  &  10502009001 & 31675  &                 A,A  &                 1,2 \\
    May 2019b & 10502007002  &  10502009001 & 32072  &                 B,B  &                 0,1 \\
    Aug 2019a & 10502010001  &  10502009001 & 19038  &             A,A,B,B  &             0,3,0,3 \\
    Aug 2019b & 10502010003  &  10502009003 & 20357  &             A,A,B,B  &             0,3,0,3 \\
    Feb 2020 & 10602003002  &  10602004002 & 32410  &             A,A,B,B  &             0,3,0,3 \\
    Feb 2020 & 10602003004  &  10602004002 & 14760  &             A,A,B,B  &             0,3,0,3 \\
    Aug 2020 & 10602005002  &  10602004005 & 14248  &             A,A,B,B  &             0,3,0,3 \\
    Aug 2020 & 10602005004  &  10602004005 & 24338  &             A,A,B,B  &             0,3,0,3 \\
    Feb 2021 & 10702305002  &  10702004002 & 23410  &             A,A,B,B  &             0,3,0,3 \\
\hline
Total & & & 527529 & & \\
\hline
\multicolumn{4}{l}{\small $^a$ list of modules where stray light is taken}\\
\multicolumn{4}{l}{\small $^b$ list of detectors to go with module list where stray light is taken}\\
\end{tabular}
\end{table}

\subsection{Stray-Light Analysis}
The SL analysis is divided into two sections: 1) the detector absorption measurement, \detabs, and 2) the RMF correction. The two effects are partially coupled but can be separated by first evaluating the detector absorption. As shown in Figure \ref{effplot} the detector absorption affects the spectrum most severely and persists all the way up to 40~keV, whereas the loss in efficiency in the spectrum due to the RMF is a much shallower function that remains approximately constant above 5~keV ($\sim98$\%). Above 4.5~keV, the detector response is well understood, but due to incorrect levels of electronic noise and pixel thresholds in the simulator used to produce the RMF there are issues below 5 keV. Since the detector absorption is the dominant component, we can get a good handle on the \detabs\ parameters even with an RMF that has low-energy issues as discussed in Section \ref{rmfcal}. The residuals that remain after the \detabs\ fitting has been performed we attribute to the RMF and correct them.

\subsubsection{Detector absorption}\label{detabs}
We built the \textsc{XSPEC} absorption model \texttt{nuabs} with cross-sections for Pt and CdZnTe created by Geant4, with the adopted photon interaction model coming from the Livermore low-energy EM model based on the evaluated photon data library, EPDL97 \cite{Cirrone2010}. We multiply \texttt{nuabs} onto the Crab model (\texttt{nuabs$\times$tbabs$\times$pow}) and fit for the Pt and CdZnTe thickness. The fit is performed using 110 data sets (there are on average 4 data sets per observation; 2 for FPMA and 2 for FPMB). We tie together the Pt and CdZnTe parameters for the same detector across all epochs, but allow the Crab spectrum to vary from epoch to epoch, only tying together the Crab parameters across detectors within the same epoch. 

\begin{figure}
\begin{center}
\includegraphics[width=0.80\textwidth]{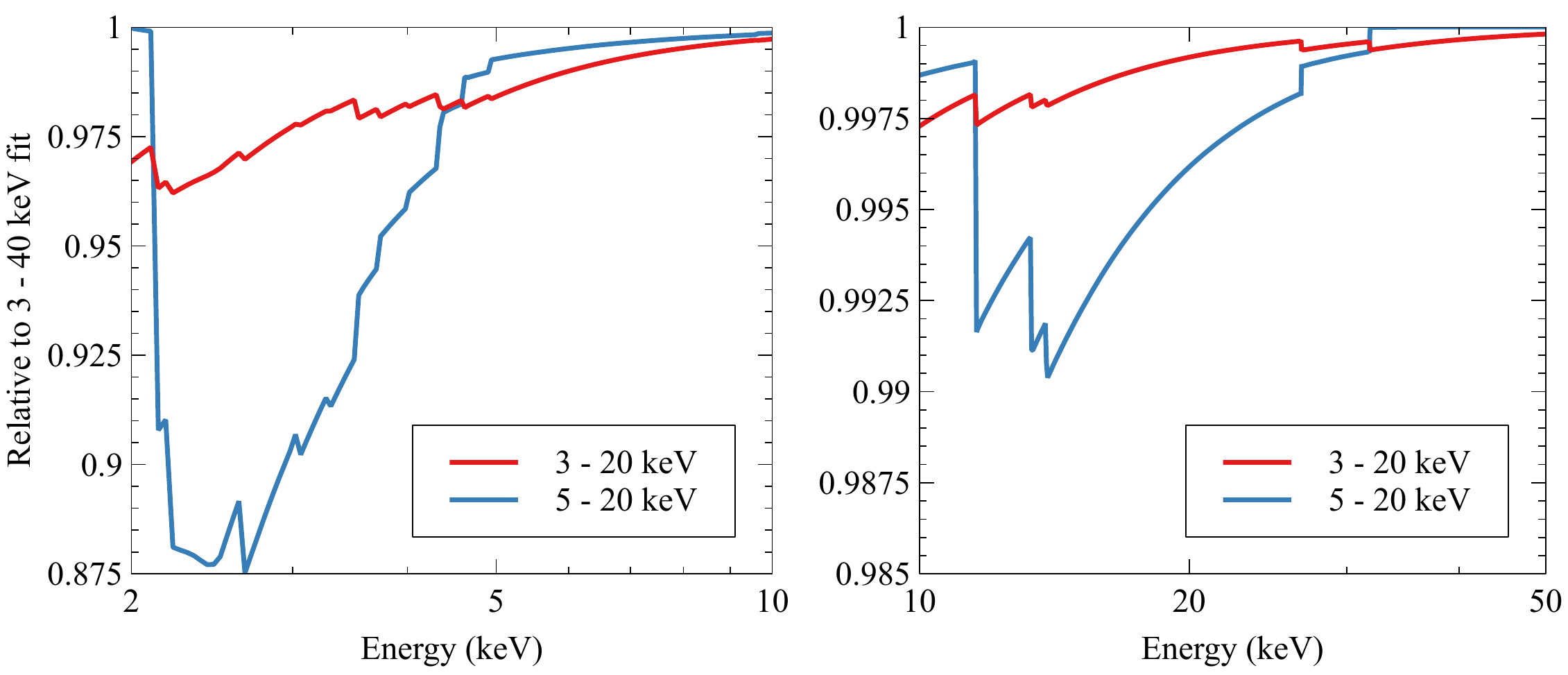}
\end{center}
\caption{Detector absorption component for Det0A for different fit energy constraints.}
\label{detabsfits}
\end{figure}

Due the fact that we are simultaneously fitting 8 detector Pt and CdZnTe parameter pairs over 20 Crab epochs, we experimented with the fitting range and the stability of the fit. We are confident about the 5--20~keV energy band, which is above where the RMF has issues and below where the background begins to affect the spectrum. The stability of the fit, however, was not good and the \detabs\ parameters took on unlikely values, which meant there were degeneracies between the Crab spectrum and \detabs. Expanding this down to 3~keV improved the stability, but the Crab spectral parameters had to be managed in order not to rail at extreme values, which again caused the \detabs\ parameters to take on unrealistic values. Expanding the energy range up to 40~keV solved the stability issues due to the long lever arm on the Crab spectrum and the forceful restriction on allowed values \detabs\ can take between 20--40~keV. We show in Figure \ref{detabsfits} the changes in the \detabs\ absorption in the 3--20~keV and 5--20~keV range with respect to 3--40~keV to illustrate the low-energy fit issue, and the larger range places solid constraints on \detabs\ despite the RMF issues.

\begin{figure}
\begin{center}
\includegraphics[width=0.80\textwidth]{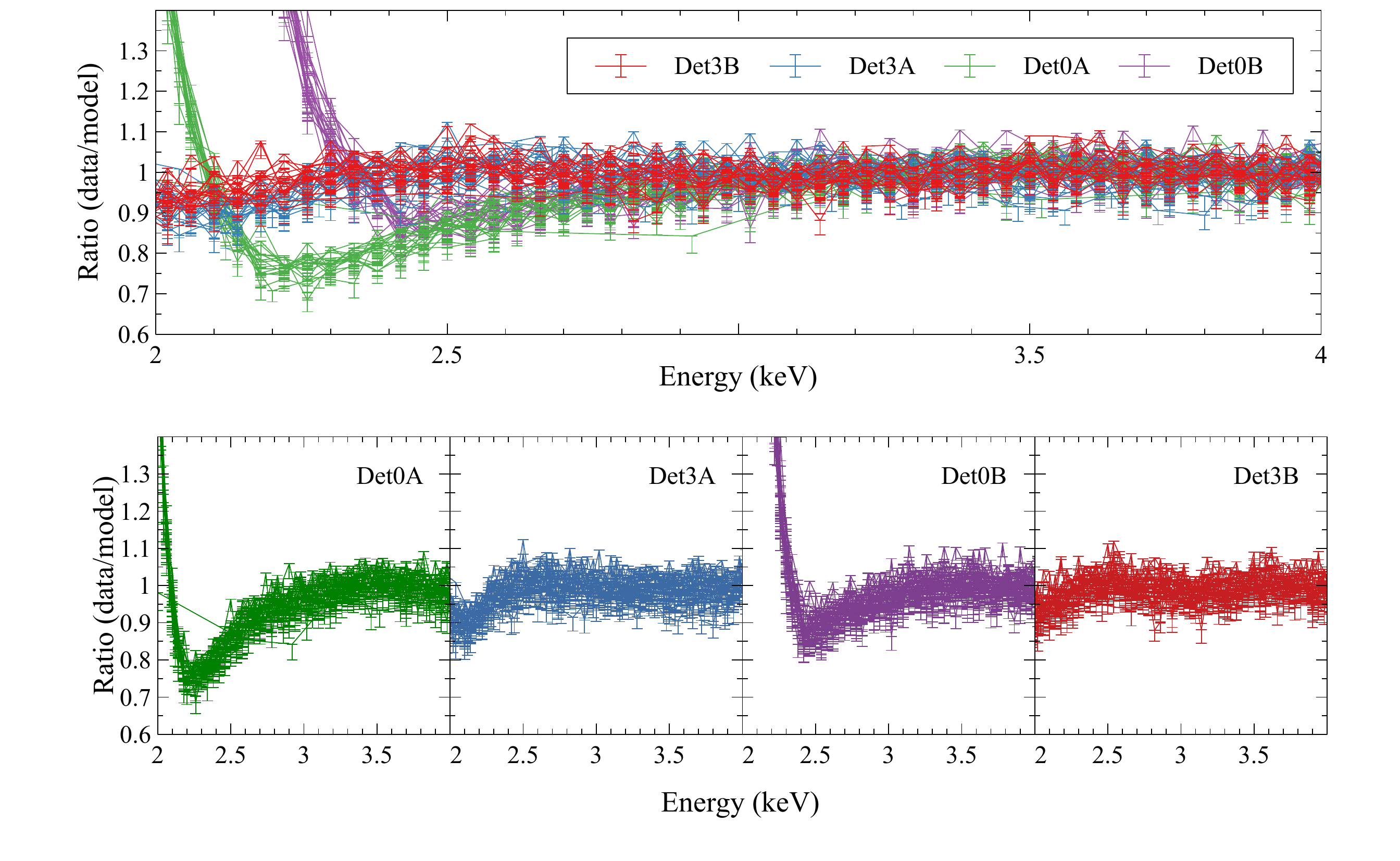}
\end{center}
\caption{Residuals for FPMA and FPMB, Det0 and Det3, after \texttt{nuabs} fitting.}
\label{lowEresidual}
\end{figure}

The key parameters for \detabs\ v004 to the 3--40~keV fit are summarized in Table \ref{nuabs}, and Figure \ref{lowEresidual} shows the remaining residuals for FPMA and FPMB, Det0 and Det3, for all observations at 2--4~keV after the fit. The residuals are independent of epoch and distinct for each detector, which supports the idea that the remaining issues are tied to the detector response.

\begin{table}
\begin{center}
\caption{\texttt{DETABS} v004 parameters.}
\begin{tabular}{c|c|c|c|c|c}
\hline
Module & Detector & Pt ($\mu$m) & 1$\sigma$ & CdZnTe ($\mu$m) & 1$\sigma$ \\
\hline
\hline
A & 0 & 0.090 & 0.001 & 0.245 & 0.008 \\
A & 1 & 0.093 & 0.004 & 0.276 & 0.026 \\
A & 2 & 0.072 & 0.005 & 0.419 & 0.030 \\
A & 3 & 0.100 & 0.001 & 0.250 & 0.008 \\
\hline
B & 0 & 0.101 & 0.001 & 0.235 & 0.008 \\
B & 1 & 0.080 & 0.003 & 0.274 & 0.020 \\
B & 2 & 0.081 & 0.012 & 0.284 & 0.070 \\
B & 3 & 0.084 & 0.001 & 0.225 & 0.008 \\
\hline
\end{tabular}
\label{nuabs}
\end{center}
\end{table}

\begin{table}
\begin{center}
\caption{Crab spectral parameters. Fit 5--20~keV}
\begin{tabular}{l|ccc}
\hline
Epoch & $\Gamma$ & Normalization &  Flux (3--10~keV) \\
 & & & ($10^{-8}$~ergs~cm$^{-2}$~s$^{-1}$) \\
 \hline
 \hline
  Oct 2015 & $  2.091 \pm  0.010 $ & $  9.32 \pm   0.06 $ & $ 1.540 \pm  0.010 $ \\
 Apr 2016 & $  2.109 \pm  0.009 $ & $  9.73 \pm   0.08 $ & $ 1.560 \pm  0.012 $ \\
 Feb 2017 & $  2.105 \pm  0.010 $ & $  9.61 \pm   0.06 $ & $ 1.552 \pm  0.010 $ \\
 Apr 2017 & $  2.101 \pm  0.010 $ & $  9.60 \pm   0.03 $ & $ 1.561 \pm  0.005 $ \\
 Jul 2017 & $  2.103 \pm  0.010 $ & $  9.63 \pm   0.07 $ & $ 1.561 \pm  0.011 $ \\
 Sep 2017 & $  2.101 \pm  0.010 $ & $  9.59 \pm   0.03 $ & $ 1.558 \pm  0.005 $ \\
 Sep 2017 & $  2.097 \pm  0.005 $ & $  9.47 \pm   0.06 $ & $ 1.549 \pm  0.010 $ \\
 Oct 2017 & $  2.102 \pm  0.010 $ & $  9.53 \pm   0.06 $ & $ 1.548 \pm  0.010 $ \\
 Nov 2017 & $  2.101 \pm  0.005 $ & $  9.65 \pm   0.05 $ & $ 1.569 \pm  0.008 $ \\
 Mar 2018 & $  2.101 \pm  0.011 $ & $  9.62 \pm   0.08 $ & $ 1.562 \pm  0.012 $ \\
 May 2018 & $  2.101 \pm  0.007 $ & $  9.63 \pm   0.05 $ & $ 1.566 \pm  0.008 $ \\
 Sep 2018 & $  2.099 \pm  0.009 $ & $  9.49 \pm   0.06 $ & $ 1.548 \pm  0.009 $ \\
 Feb 2018 & $  2.107 \pm  0.009 $ & $  9.61 \pm   0.03 $ & $ 1.545 \pm  0.005 $ \\
 Mar 2019 & $  2.102 \pm  0.020 $ & $  9.52 \pm   0.13 $ & $ 1.546 \pm  0.021 $ \\
May 2019a & $  2.105 \pm  0.012 $ & $  9.65 \pm   0.08 $ & $ 1.558 \pm  0.013 $ \\
May 2019b & $  2.092 \pm  0.010 $ & $  9.39 \pm   0.06 $ & $ 1.548 \pm  0.010 $ \\
Aug 2019a & $  2.099 \pm  0.009 $ & $  9.49 \pm   0.06 $ & $ 1.549 \pm  0.010 $ \\
Aug 2019b & $  2.101 \pm  0.006 $ & $  9.64 \pm   0.05 $ & $ 1.567 \pm  0.008 $ \\
 Feb 2020 & $  2.107 \pm  0.006 $ & $  9.76 \pm   0.06 $ & $ 1.571 \pm  0.009 $ \\
 Aug 2020 & $  2.098 \pm  0.007 $ & $  9.55 \pm   0.03 $ & $ 1.560 \pm  0.005 $ \\
\hline
Epoch average $\dagger$ & $2.103 \pm 0.002$ & $9.65 \pm 0.01 $ & $1.56 \pm 0.002$ \\
Epoch average $\ddagger$ & $2.103 \pm 0.002$ & $9.73 \pm 0.01 $ & $1.57 \pm 0.002$ \\
\hline
\textbf{Canonical} & $2.103 \pm 0.002$ & $9.69 \pm 0.01 $ & $1.57 \pm 0.002$ \\
\hline
\hline
\multicolumn{4}{l}{$\dagger$ Simultaneous fit to all 110 SL data sets} \\
\multicolumn{4}{l}{$\ddagger$ Simultaneous fit to detector combined data} \\
\end{tabular}
\label{crabepoch}
\end{center}
\end{table}

\subsubsection{Epoch averaged Crab SL spectrum}\label{avgcrab}

When computing the corrections to the RMF (Section \ref{rmfcal}) and the vignetting function (Section \ref{vignetting}) we need to produce a canonical Crab model to calibrate against. Ideally, we would want to do this epoch-by-epoch with SL observations matched to every focused observation. We unfortunately do not have matched SL and focused observations (Table \ref{SLobslog} and Table \ref{focusobslog}) for all epochs. In addition, the new calibration method described in Section \ref{newmethod} also necessarily mixes together different epochs to produce spectra binned by off-axis angle. Finally, to produce enough signal-to-noise to reduce the statistical errors to the $\sim$2\% we have to combine all of the SL data for each detector, thereby also mixing epochs.


We include the variations in the Crab spectrum itself as a source of systematic noise, and we can evaluate the approximate magnitude of this systematic error by looking at variations within the \nustar\ data and comparing this to monitoring instruments such as Swift/BAT, which will be done in Section \ref{result}. Here, we assume that both the RMF and the Crab are stable in time.

To produce the canonical model, we fit the SL data from all epochs over the energy range between 5--20~keV, because in this range the responses are best known and minimally impacted by the low-energy RMF issues (to be calibrated out the next section) and background inaccuracies. We performed this epoch-average fit in two ways: 1) fitting all of the 110 SL data sets simultaneously to the same Crab model, and 2) first combining all data and background into detector-specific spectra then fit to the same Crab model. We load the \detabs\ v004 parameters from Table \ref{nuabs} and keep them frozen in the \texttt{nuabs} model. The results from the two methods are shown in Table \ref{crabepoch}, and they are consistent within errors, so we adopt the average of these two results, $\Gamma = 2.103 \pm 0.001$ and N$ = 9.69\pm 0.02$~keV$^{-1}$ cm$^{-2}$ s$^{-1}$ at 1 keV, and hereafter define to the canonical Crab model to be: $\Gamma \equiv 2.103$ and $N \equiv 9.69$.



\subsubsection{RMF calibration}\label{rmfcal}

The readout architecture for the \nustar\ detectors directly measures the charge collected by each pixel. The resulting signal is sent to a ``fast" trigger chain, which has a shaping time of a few $\mu$s. If the signal exceeds the trigger criteria, then the event is read out. Two main parameters involved in the readout of the fast chain are: the trigger threshold itself ($\mu$) and the noise on the shaping amplifier ($\sigma$) \cite{grefenstette_2018_threshold}. The trigger thresholds are typically set to be $\sim$ 2~keV, though this is a software-programmable value that changes the average threshold for each detector, and pixel-to-pixel component variations can lead to effective ``gain" differences on the fast chain that affect the apparent value of $\mu$. $\sigma$ may also vary detector-by-detector.

In our comparison of the Crab model with the RMF, we found that Det0 on FPMA and FPMB are both close to the nominal 2 keV threshold. However, for the other three detectors we measure a lower threshold. In fact, for FPMB Det1 and Det3, the threshold is actually below the nominal ``PI channel 0". This is due to an artificial floor in the definition of the \nustar\ PI channel scale, which starts at 1.6 keV. In principle, this means that we could adjust the definition of the PI channels down to enable more low-energy analysis of the \nustar\ data. However, this is beyond the scope of the current calibration effort as it would likely result in substantial confusion in the community.

In addition to the differences in the threshold value, we also noted that the measured drop off in quantum efficiency was markedly broader than what was captured in the simulated RMFs. This can  be explained by an unmodeled source of noise in the fast electronics chain that was not accounted for (e.g., the value of $\sigma$ was incorrect). We note that this noise term is completed unrelated to any energy resolution effects and \textit{only} affects the shape of the quantum efficiency curve at low energies.

\begin{table}[ht]
\begin{center}
\caption{RMF threshold parameters.}
\begin{tabular}{c|c|c}
\hline
Module & Detector & $\mu$ (keV) \\
\hline
\hline
A & 0 & 2.0 \\
A & 1 & 1.83  \\
A & 2 & 1.88 \\
A & 3 & 1.7  \\
\hline
B & 0 & 2.1 \\
B & 1 & 1.42 \\
B & 2 & 1.68 \\
B & 3 & 1.45 \\
\hline
\end{tabular}
\label{thresh}
\end{center}
\end{table}

In the previous version of the GEANT4 code that generated the RMF, the threshold values for all pixels was assumed to have a single value (2 keV) and the $\sigma$ term was assumed to be negligible. This resulted in an overly ``sharp" cutoff in the RMF at 2 keV. Because a large fraction of the events in \nustar\ result from ``sharing" charge between neighboring pixels, the impact of improperly calibrating the RMF threshold could become apparent at roughly 2x the trigger level (so, 4 keV).

\begin{figure*}
\begin{center}
\includegraphics[width=0.45\textwidth]{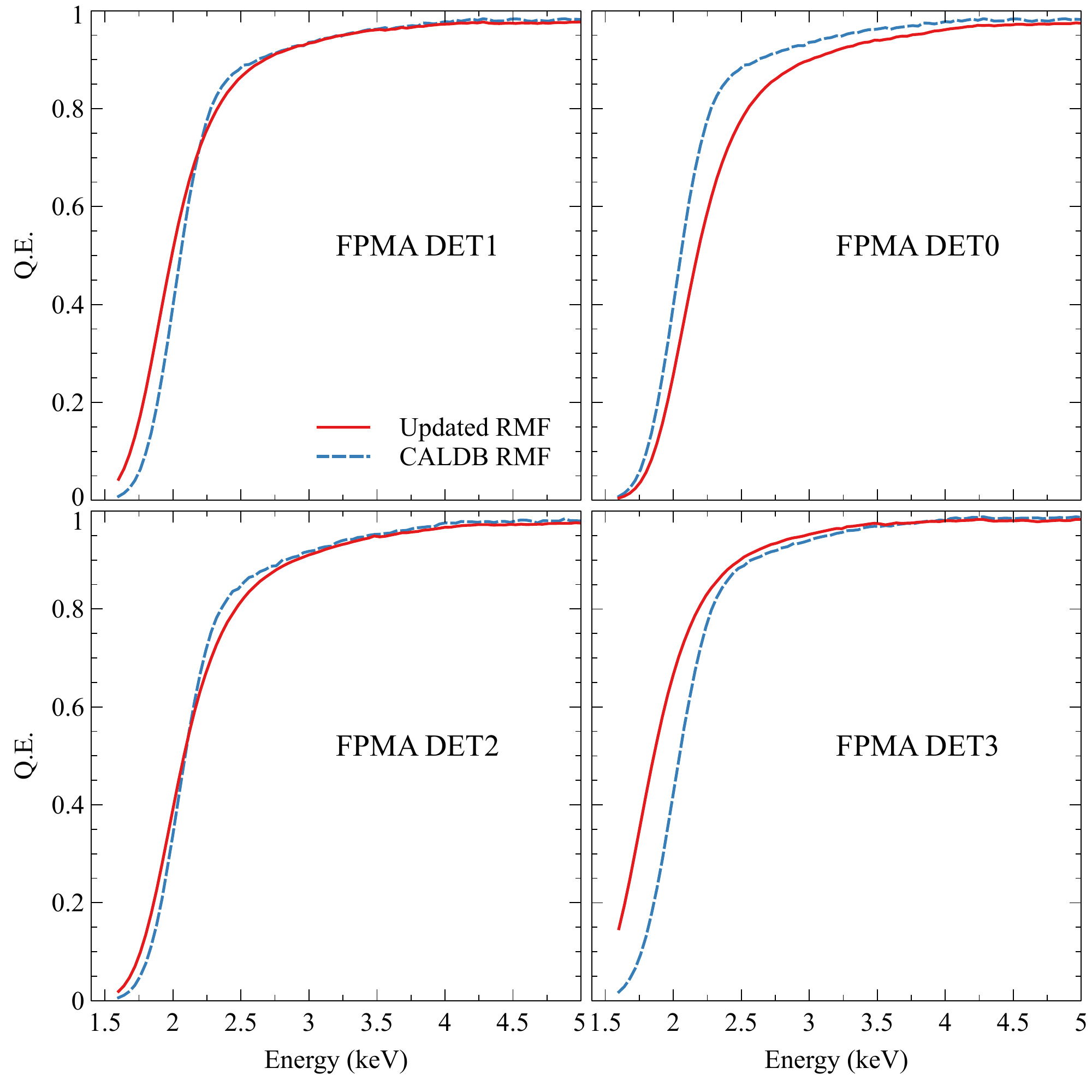}
\includegraphics[width=0.45\textwidth]{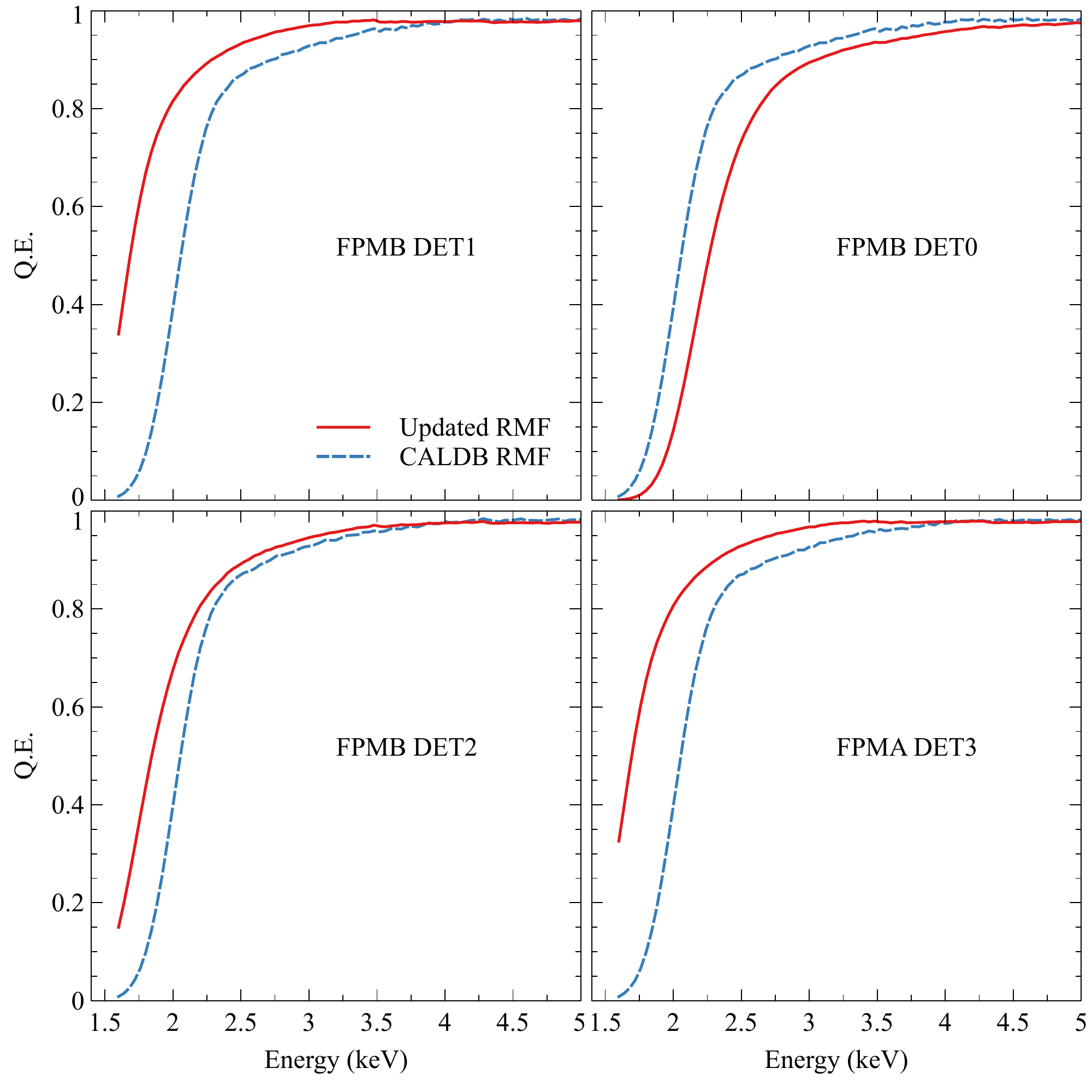} \\
\includegraphics[width=0.45\textwidth]{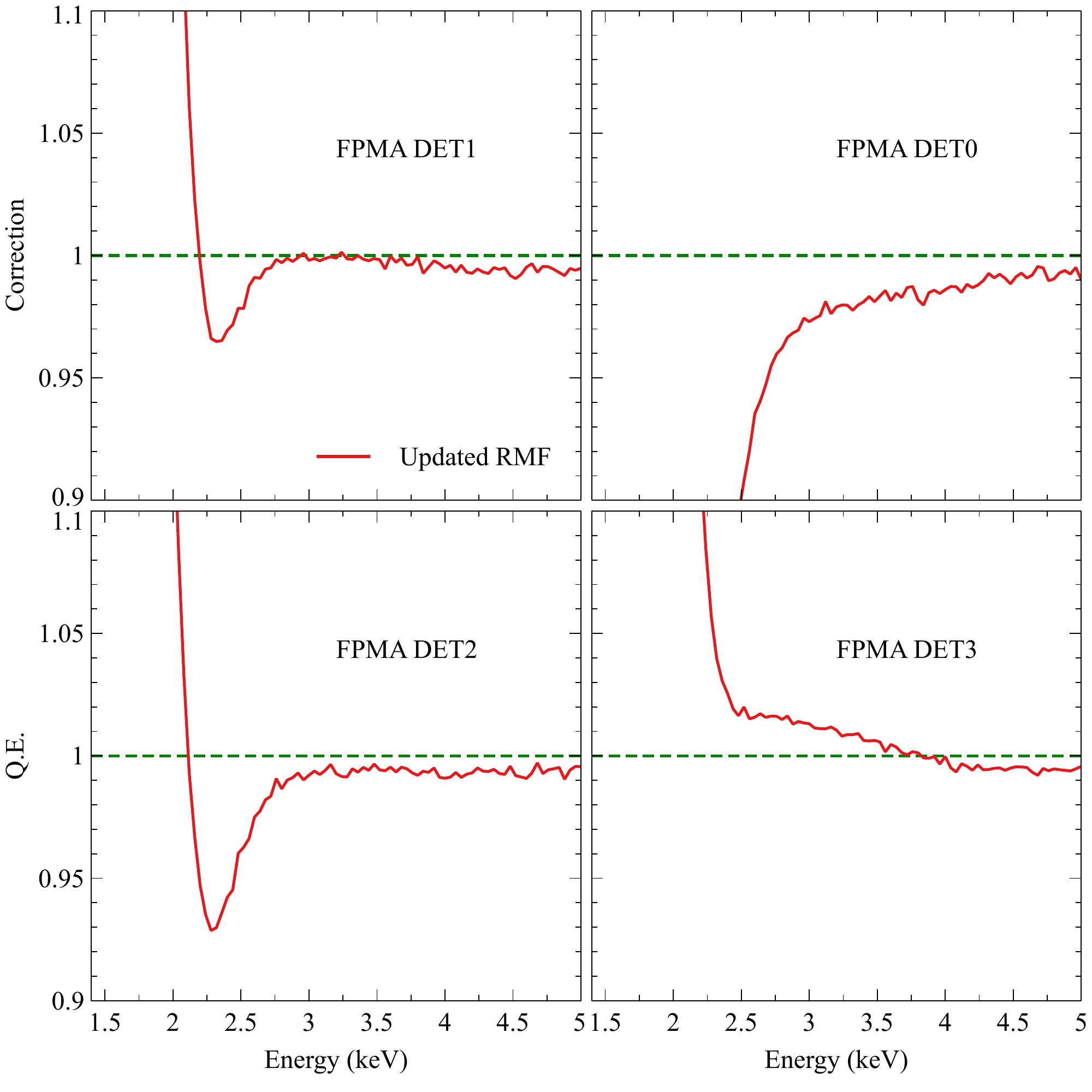}
\includegraphics[width=0.45\textwidth]{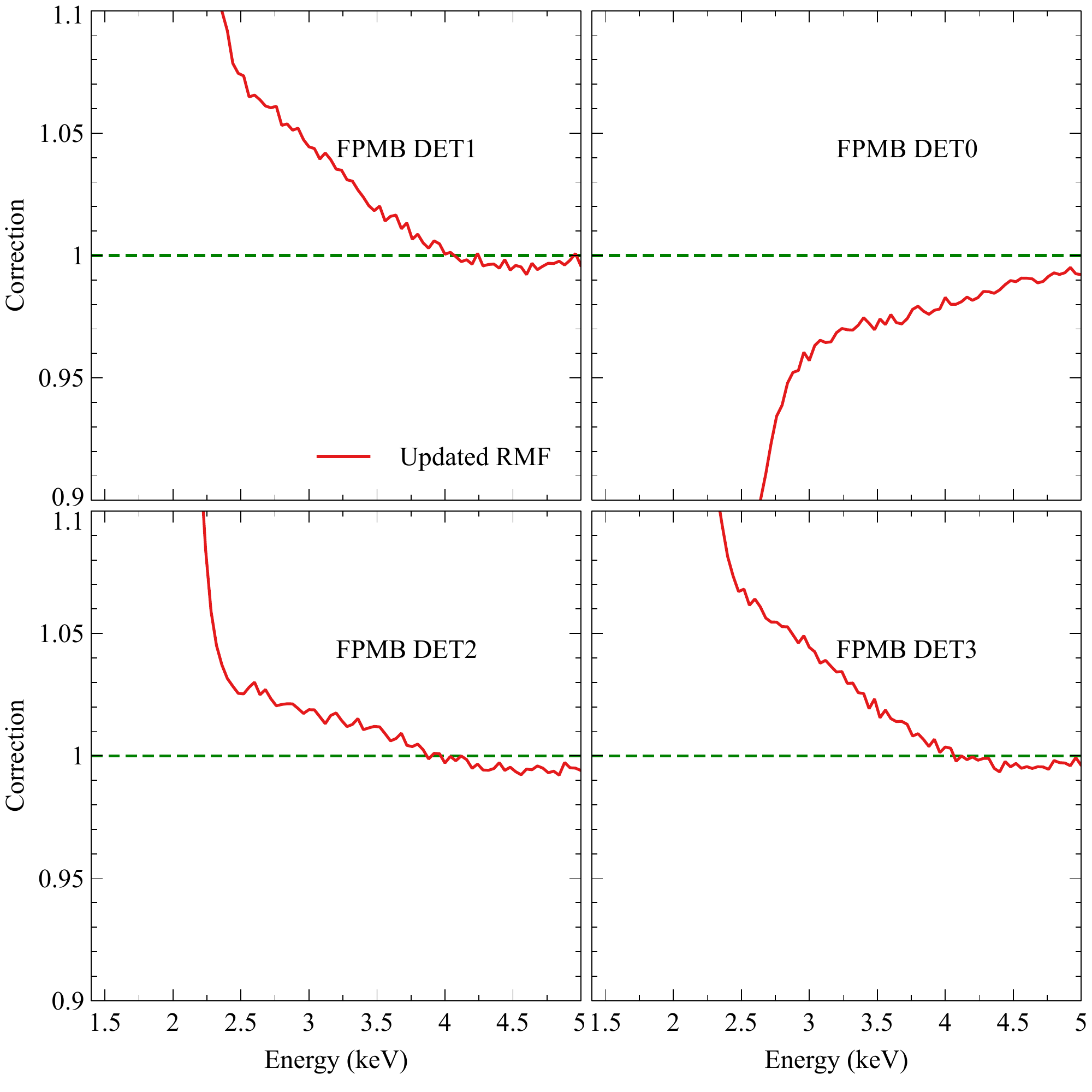}
\end{center}
\caption{Top Panel: Comparison of the simulated quantum efficiency in the v3.0 RMFs for FPMA (left) and FPMB (right) for the CALDB RMFs (blue, dashed) compared with the RMFs with the corrected $\mu$ and $\sigma$ values. Bottom Panel: The effect of implementing the new RMF, including the characteristic ``swoop" residual seen in Figure \ref{lowEresidual}.
}
\label{rmf_comparison}
\end{figure*}

We updated the RMF code to properly account for the coarse measurements of $\mu$ and $\sigma$ made using the ``nearest neighbor" analysis method\cite{grefenstette_2018_threshold}. For all detectors, the noise term was consistent with 0.7 keV, while the updated $\mu$ values are shown in Table \ref{thresh}. In addition to adjusting the threshold, we also increased the number of incident photons from 1.44M to 14.4M to reduce the impact of numerical statistical fluctuations in the QE curve to roughly 0.35\%. A comparison of the 2010v002 RMFs and the updated RMFs v3.0 from the GEANT4 simulation are shown in Figure \ref{rmf_comparison}.

\begin{figure}
\begin{center}
\includegraphics[width=0.32\textwidth]{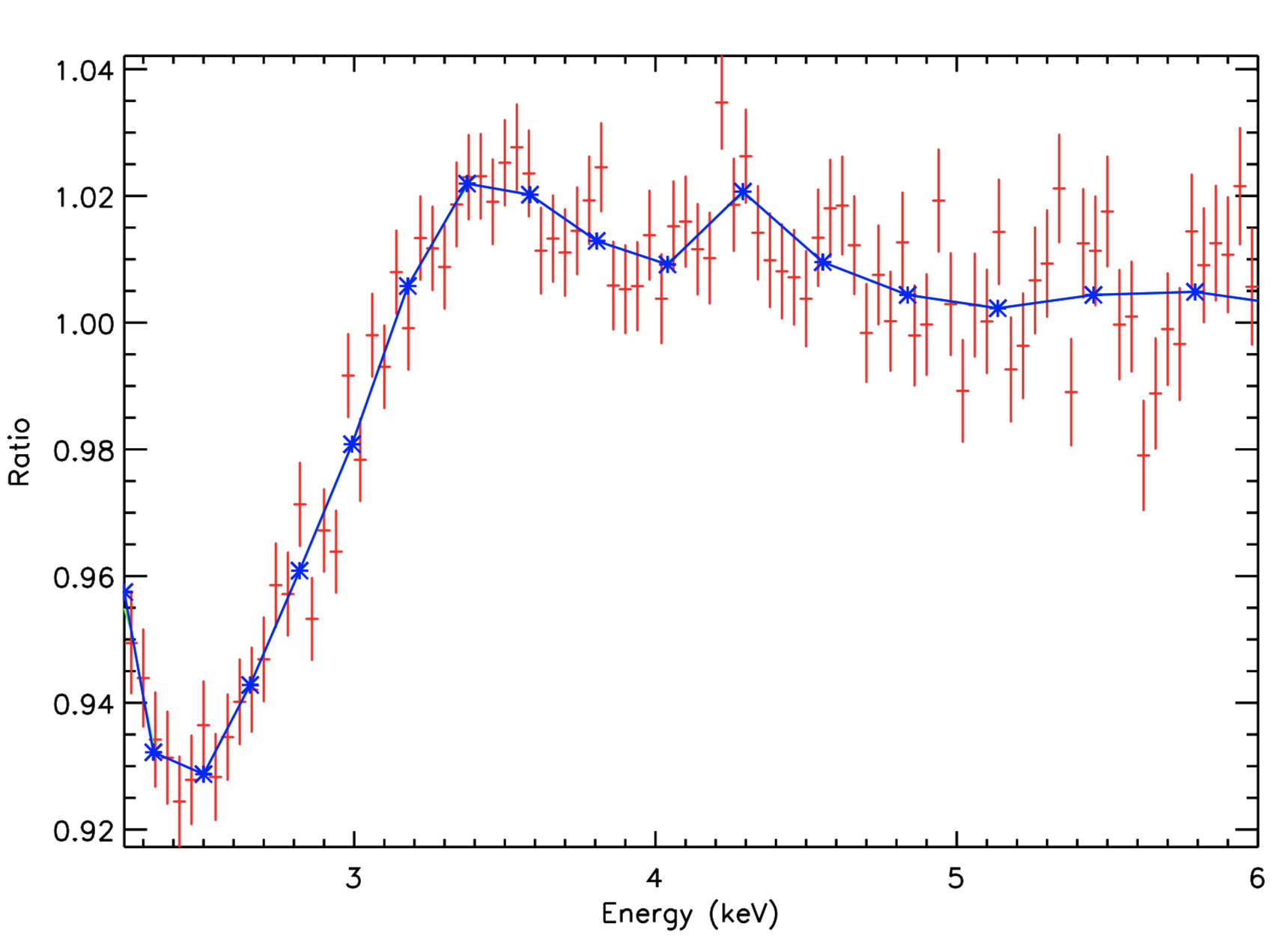}
\includegraphics[width=0.32\textwidth]{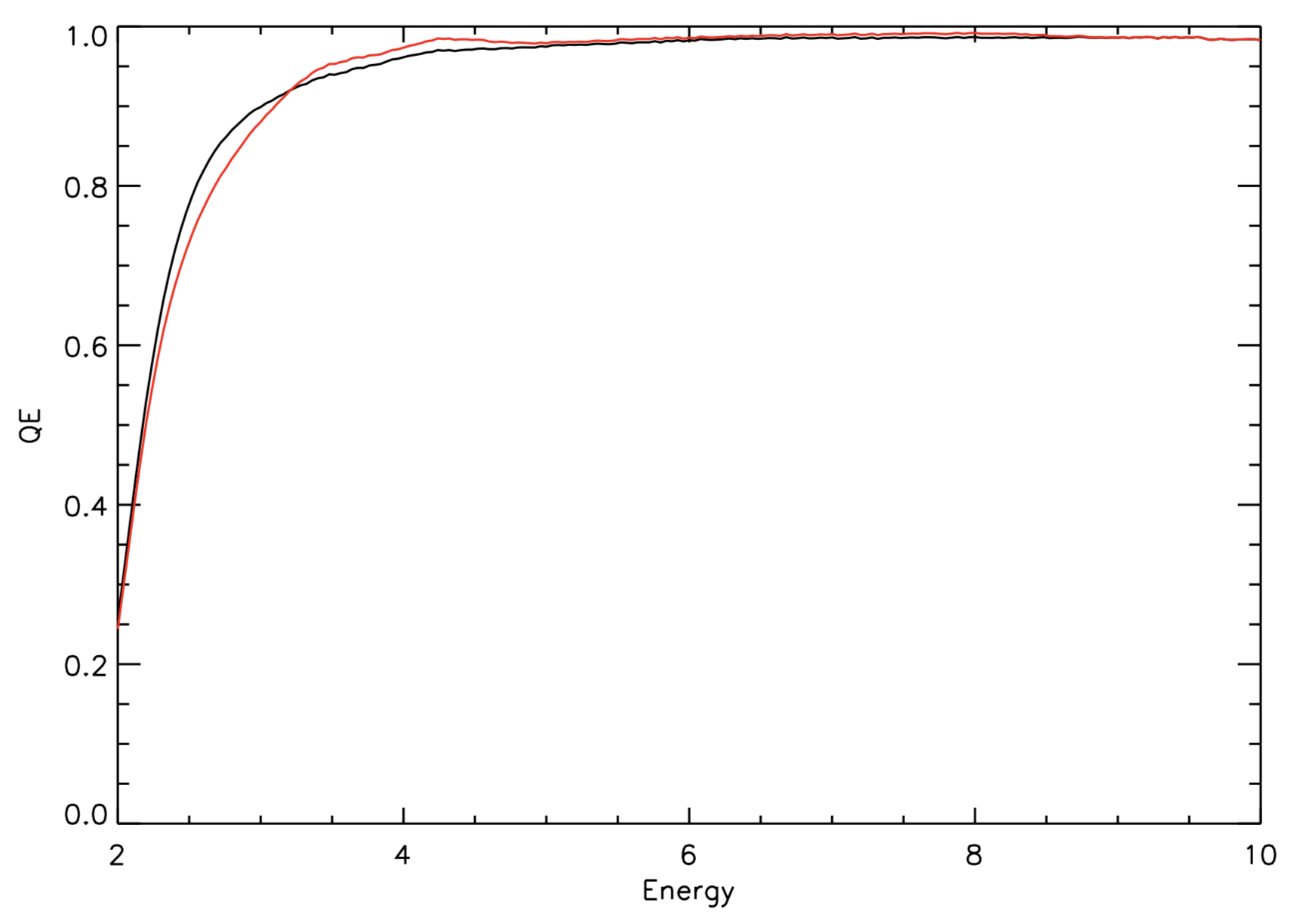}
\includegraphics[width=0.32\textwidth]{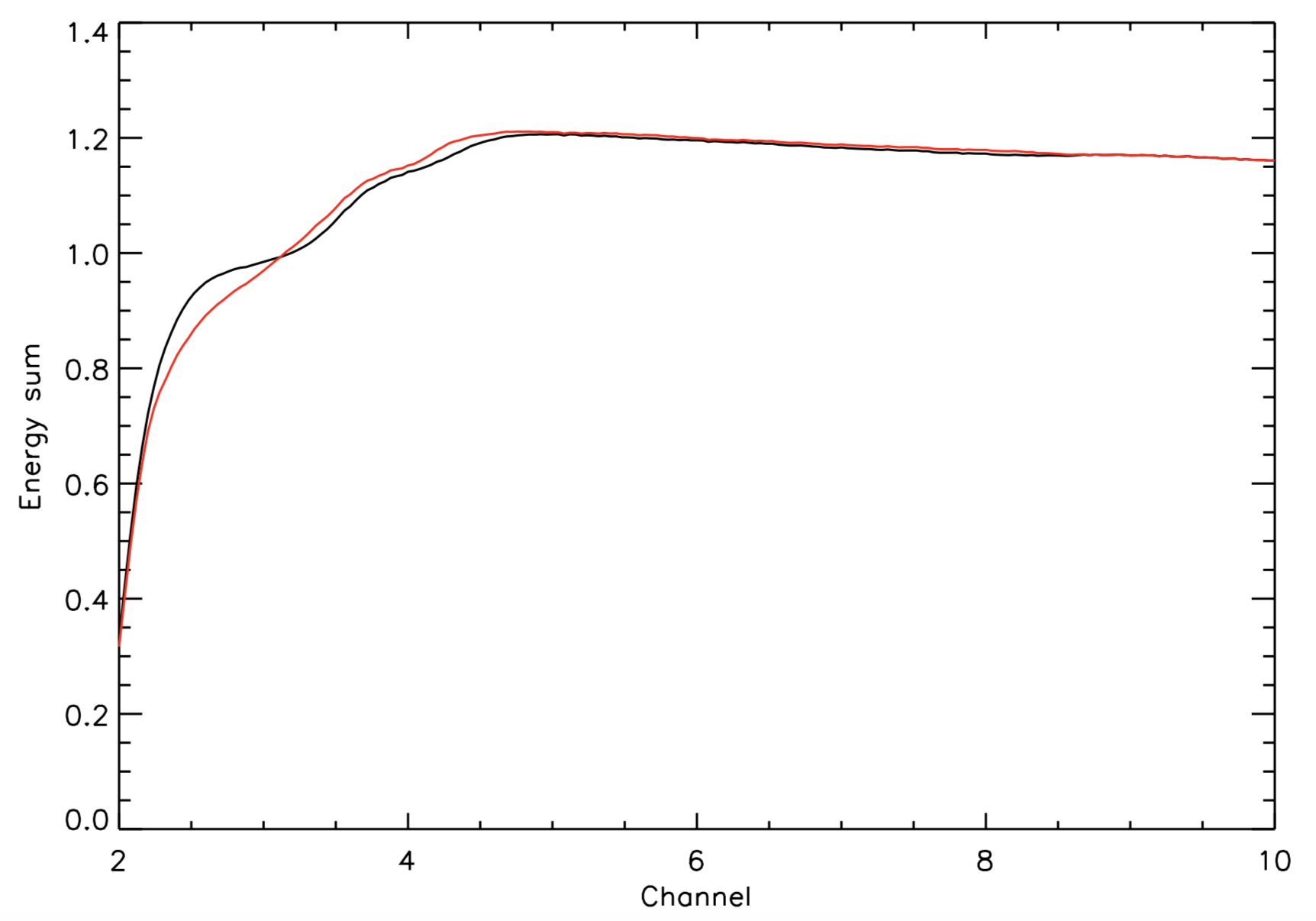}
\end{center}
\caption{Left: Residuals to the canonical model before correction, over plotted (in blue) with the correction folded through the response and where stars donate the grid points of the fit. Middle: RMF v3.0 (black) and RMF v3.1 (red) summed along the channel axis. Right: RMF v3.0 (black) and RMF v3.1 (red) summed along the energy axis.}
\label{rmffit}
\end{figure}

The SL data were reduced using this intermediate RMF v3.0, and to correct the features seen in Figure \ref{lowEresidual}, we applied a correction function on the RMF v3.0 input channel. This effectively means the correction is redistributed in energy and the output efficiency is a superposition of corrections from several channels. Figure \ref{rmffit} shows an example fit for Det0A and illustrates that the corrected RMF v3.1 has more structured variations at the 2--4~keV turnover than the simulated RMF v3.0.

\section{Effective Area calibration}\label{vignetting}
\subsection{Observations and Data Reductions}\label{reduction}

\begin{figure}
\begin{center}
\includegraphics[width=0.70\textwidth]{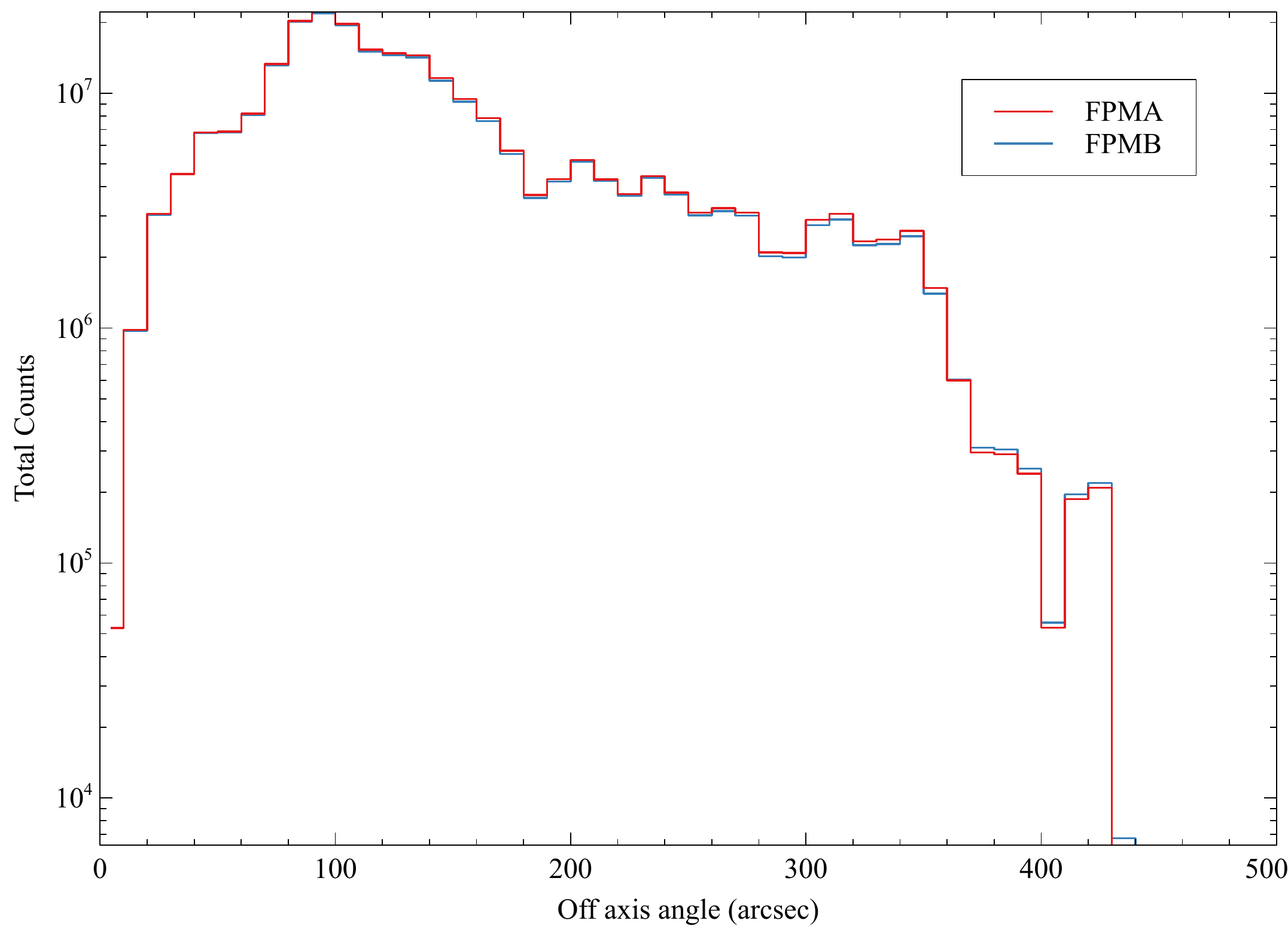}
\end{center}
\caption{Histogram of the count distribution between 3-78~keV as a function of off-axis angle.}
\label{statistic}
\end{figure}

The on-going \textit{NuSTAR} Crab calibration campaign measures the Crab spectrum at various off-axis angles. This has to date resulted in 71 individual observations listed in Table \ref{focusobslog} with a total, single FPM exposure time of 233~ks. Figure \ref{statistic} shows the distribution of all counts within the 3--78~keV band as a function of off-axis angle for FPMA and FPMB. The counts peak at $\sim$1.5\am, which is the default \nustar\ `on-axis' position (moving closer to the optical axis causes the source to fall into detector gaps), but overall there is more than $10^6$ counts per off-axis bin.

For the data reduction, we treat the Crab as a point source despite the angular extent of the Crab, which is approximately $\sim$120\as$\times100$\as, since the center of the nebula where the pulsar resides dominates the emission. The difference in the effective area response between a point source extraction centered on the pulsar and an extended source extraction is only $\sim$1-2\% and therefore an acceptable error. Buy most importantly, by treating the Crab as a point source it allows us to use the built-in pipeline corrections for the aperture stop, ghost ray, and PSF corrections (for details on these components see Madsen et al. (2015)\cite{Madsen2015a}and (2017)\cite{Madsen2017a,Madsen2017b}). These corrections are not applicable to extended source responses, and since for certain off-axis angels they can be larger than the error introduced by assuming the Crab is a point source, they are important to include. We extract counts from a 200\as\ region, which includes $\sim$95\% of all photons in the source extraction region. The 95\% are corrected for in the PSF.

We run \texttt{nupipeline} with default settings, but include the \texttt{statusexpr} keyword (STATUS == b0000xx000 xxxx000), which allows good events rejected due to high countrates to be included. In addition to default settings, we also applied the temperature dependent MLI correction for the identified subset marked in Table \ref{focusobslog}, and we excluded some FPMA or B observations if the pulsar of the Crab fell into a detector gap. Since much of the flux comes from the pulsar, which has a significantly harder spectrum than the PWN\cite{Madsen2015b}, both simulations and observations show that loss of this component will distort the spectrum and make it unsuitable for calibration. After the pipeline has run, we find the GTIs that subdivides each observation into the respective off-axis angle bins. The source off-axis angle is the angular distance between the optical axis and the source center. We calculate the off-axis angle as a function of time from the '\_det1.fits' file produced by \texttt{nupipeline}, which defines for each time bin where the optical axis was in coordinates of the sky. The generated GTI files are input into \texttt{nuproducts} to split the observations into the angle specific spectra. We combine the GTI sorted spectra and their backgrounds into the final off-axis angle bin using the FTOOL \texttt{addascaspec} and combine the RMF using \texttt{addrmf}.

\begin{table}
\centering
\caption{Focused Crab Observing Log}\label{focusobslog}
\begin{tabular}{lcc|lcc}
\hline
Obsid &  mean off-axis angle & exposure time & Obsid &  mean off-axis angle & exposure time\\
 &  (arcmin) &  (s) &  &  (arcmin)  & (s)\\
\hline
\hline
         10013021002  & 3.6  &    2275  &         10013022002  & 1.5  &    2592 \\
         10013022004  & 1.5  &    2347  &         10013022006  & 1.4  &    2587 \\
         10013023002  & 4.1  &    2102  &         10013024002  & 4.7  &    2258 \\
         10013025002  & 5.0  &    1235  &         10013025004  & 5.0  &    1161 \\
         10013025006  & 4.8  &    1593  &         10013026002  & 5.5  &    2540 \\
         10013026004  & 5.7  &    1162  &         10013027002  & 5.8  &    1182 \\
         10013027004  & 5.6  &    1105  &         10013028002  & 6.5  &    1601 \\
         10013028004  & 6.7  &    1254  &         10013029001  & 4.4  &    2909 \\
         10013030001  & 4.4  &    3023  &         10013031002  & 1.9  &    2507 \\
         10013032002  & 1.3  &    2595  &         10013033002$^*$  & 1.5  &    1383 \\
         10013033004$\ddagger$  & 1.3  &    1269  &         10013034002  & 1.2  &     988 \\
         10013034004$^+$  & 0.8  &    5720  &         10013034005$^+$  & 0.9  &    5968 \\
         10013035002  & 5.4  &    9401  &         10013036002  & 5.8  &     179 \\
         10013037002  & 2.0  &    2679  &         10013037004  & 2.9  &    2796 \\
         10013037006  & 2.9  &    2799  &         10013037008  & 3.1  &    2814 \\
         10013038002  & 3.9  &    3084  &         10013038004  & 3.9  &    2217 \\
         10013038006  & 4.0  &     266  &         10013038008  & 4.0  &    2231 \\
         10013039002  & 4.1  &     590  &         10013039003  & 4.2  &     583 \\
         80001022002  & 1.5  &    3917  &         10002001002  & 1.6  &    2608 \\
         10002001004  & 1.6  &    2386  &         10002001006  & 1.4  &   14263 \\
         10002001008  & 1.6  &    4941  &         10002001009  & 1.3  &    5198 \\
         10202001002$^*$  & 1.1  &    1454  &         10202001004  & 2.0  &    9025 \\
         10202001006  & 1.7  &    4307  &         10202001007  & 2.9  &    4104 \\
         10302001002  & 2.5  &    4300  &         10302001004  & 1.2  &    7547 \\
         10302001006$^*$  & 1.0  &    3729  &         10402001002$^*$  & 1.1  &    9436 \\
         10402001004  & 1.4  &    3312  &         10402001006  & 1.4  &    4757 \\
         10402001008  & 1.3  &    5657  &         10402001010  & 1.9  &    1133 \\
         10402001012  & 1.8  &    1270  &         10402001014  & 1.8  &    1284 \\
         10402001016  & 1.8  &    1169  &         10502001002  & 2.8  &    2676 \\
         10502001004  & 2.3  &    2022  &         10502001006$\dagger$  & 2.2  &    3949 \\
         10502001008$\dagger$  & 1.9  &    2417  &         10502001010  & 2.1  &     432 \\
         10502001011  & 4.0  &    9553  &         10502001013  & 2.0  &    7665 \\
         10502001015  & 2.3  &    7492  &         10602002002  & 2.2  &    3567 \\
         10602002004  & 2.1  &    3590  &         10602002006  & 1.9  &    2209 \\
         10602002008  & 2.0  &    2514  &         10702303002  & 1.9  &    2957 \\
         10702303004  & 2.0  &    3150  &                      &      &   \\
\hline
\multicolumn{5}{l}{Total: } & 233020 s\\
\hline
\multicolumn{6}{l}{$\dagger$ MLI temperature correction} \\
\multicolumn{6}{l}{$\ddagger$ Not included for FPMA} \\
\multicolumn{6}{l}{$^+$ Not included for FPMB} \\
\multicolumn{6}{l}{$^*$ Not used for either but included in tests} \\

\end{tabular}
\end{table}

\subsubsection{Background}
Obtaining backgrounds for the Crab observations presented us with some challenges. Background starts to affect the spectrum at $\sim 55$~keV, which at these energies is mostly internal detector background. However, because of the Crab's brightness and the extend of the PSF, we can not extract a local background from the same observation. Along with the stray-light observations mentioned above, we take blank backgrounds a few degrees from the Crab. There are multiple observations, and we combine the best of them into a master background. Also, since the Crab is either observed at sky Position Angle (roll of the Field of View) of PA=150 deg or PA=330 deg, we must combine the backgrounds separately for the two PA to directly apply the extraction region from the source image to the background. 

\begin{figure}
\begin{center}
\includegraphics[width=0.90\textwidth]{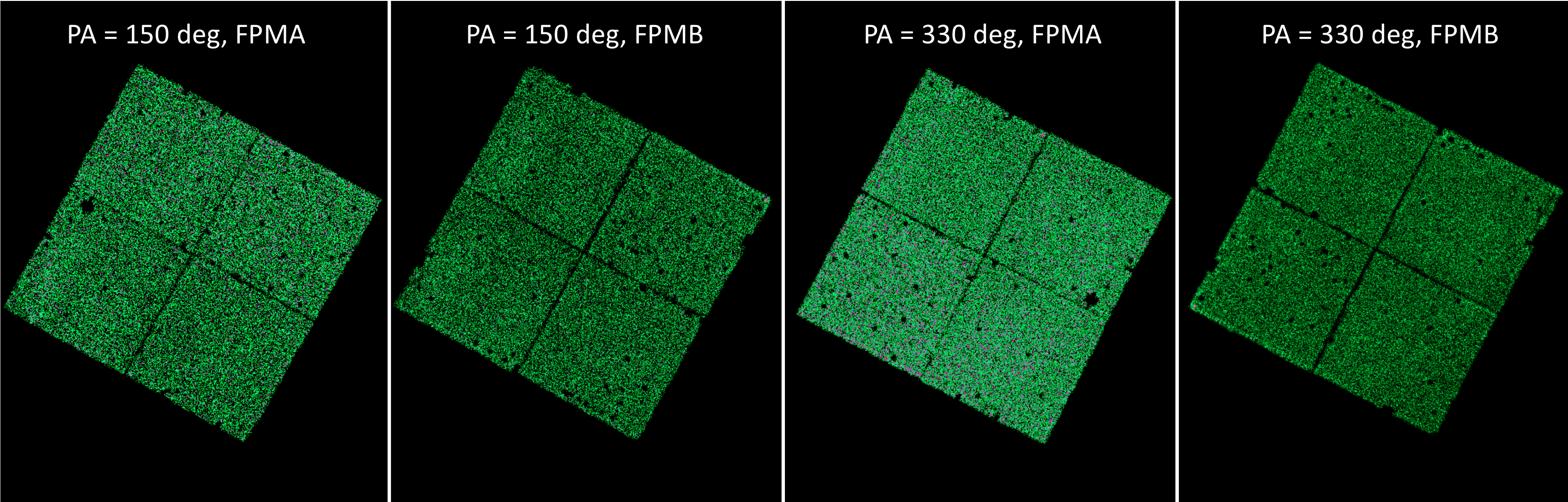}
\end{center}
\caption{Master backgrounds for PA=150 deg and PA=330 deg for FPMA and FPMB.}
\label{masterbkg}
\end{figure}

Because the projection of the detector on the sky is never the same for any observation, we have to remove the spacecraft jitter from the background observations as well before stacking them. This is a non-standard procedure, which involves correcting files that track the center of the detector on the sky to produce a jitter-free image. In this manner, we combine together several jitter-corrected background observations listed in Table \ref{background} into two master backgrounds with over 100ks exposure as shown in Figure \ref{masterbkg}. 

By definition when filtering on off-axis angle, the source position on the detector is stable and virtually jitter-free (since the spacecraft/mast jitter is what causes the off-axis angle variations), and we can therefore extract the background from the same location on the detector where the source falls, ensuring that we are sampling the correct instrumental background. 

We make a note here that the background treatment is the only step in the data preparation that does not involve the regular NUSTARDAS and FTOOLS. Less precise background can be obtained by extracting from individual background fields closest in time to the observation as was done for the stray-light analysis.

\begin{table}
\centering
\caption{Backgrounds For Focused Observations.}\label{background}
\begin{tabular}{cc|cc}
\hline
Obsid  & exposure time & Obsid & expsoure time \\
For PA = 150 deg & (s) & For PA = 330 deg  & (s) \\
\hline
\hline
10311002002 & 16048 & 10210002002 & 20567 \\ 
10311002006 & 14898 & 10402006002 & 17427 \\
10311002008 & 18633 & 10502004002 & 14418 \\
10402005002 & 26826 & 10502004004 & 15210 \\
10602004005 & 28207 & 10602004002 & 47886 \\
& & 10602004003 & 38184 \\
& & 10702004002 & 39740 \\
\hline
Total: & 104612 & Total: & 193432\\
\hline
\end{tabular}
\end{table}

\subsection{Fitting procedure}\label{focuscrabfit}
In section \ref{avgcrab} we found from the combined fit to all stray-light observations that $\Gamma = 2.103 \pm 0.001$ and N$ = 9.69\pm 0.02$~keV$^{-1}$ cm$^{-2}$ s$^{-1}$ at 1 keV, which we will refer to  as the canonical Crab model. We fold this model through the response functions to obtain a model count spectrum for a fixed off-axis angle, $\theta$,
\begin{eqnarray} 
C(PI)_\mathrm{model} = \mathrm{RMF}(PI,E) \mathrm{ARF}(E) Corr(E)\,,
\end{eqnarray}
where $\mathrm{ARF}(E)$ contains all the effective area corrections discussed in Section \ref{corrfunc}, and $Corr(E)$ is the correction function we wish to find. We compare this model count spectrum to the reference spectrum and use least squares fitting of a piece-wise linear function to find the array elements of $Corr(E)$. We do this for all $\theta$ (from 0\am\ to 7\am\ in steps of 10\as) to obtain $Corr(\theta,E)$.

This forward folding approach is necessary because the RMF is a non-diagonal matrix and cannot be inverted. This also means that the solution to $Corr(E,\theta)$ is not necessarily unique and care must be taken to avoid non-physical solutions. This may occur when the RMF due to energy redistribution compensates for a disproportionately deep sharp trough or high peak with with an opposite extreme. This issue may be mitigated by choosing sensible binning and sufficient statistics in the grid of fitting points, and we employ a logarithmic binning in energy that varies depending on the fidelity of the spectrum.

\begin{figure}
\begin{center}
\includegraphics[width=0.45\textwidth]{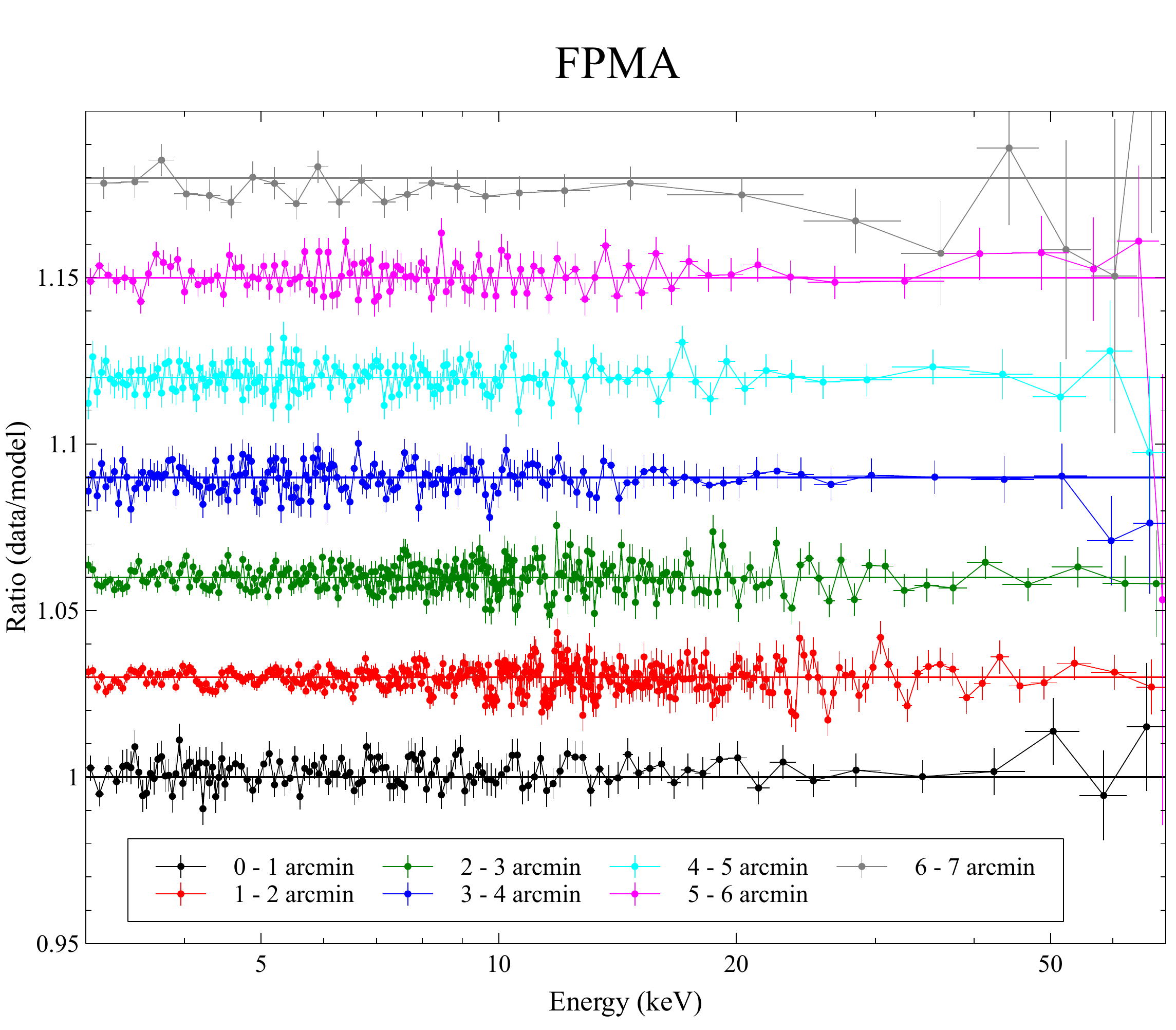}
\includegraphics[width=0.45\textwidth]{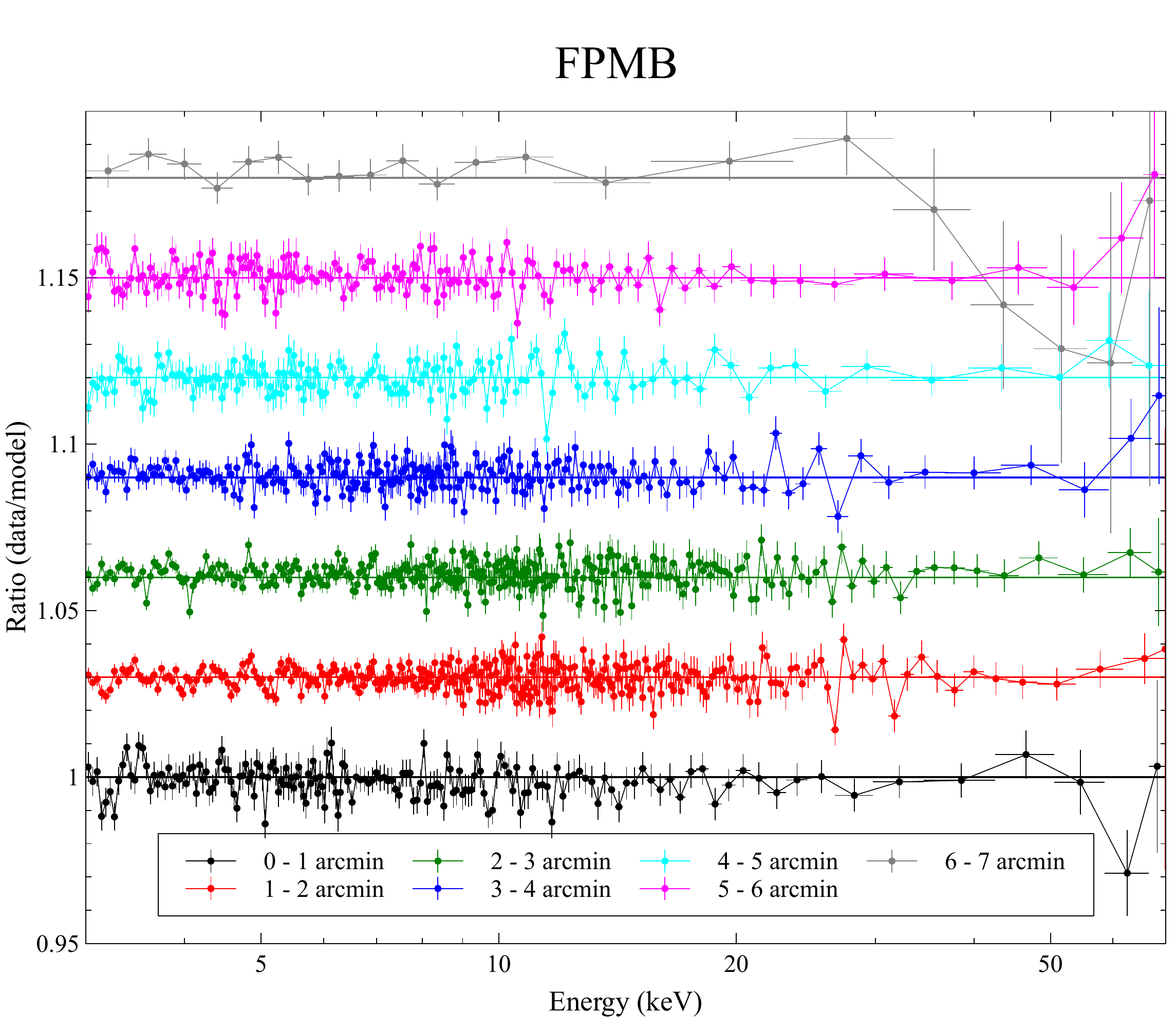}
\includegraphics[width=0.45\textwidth]{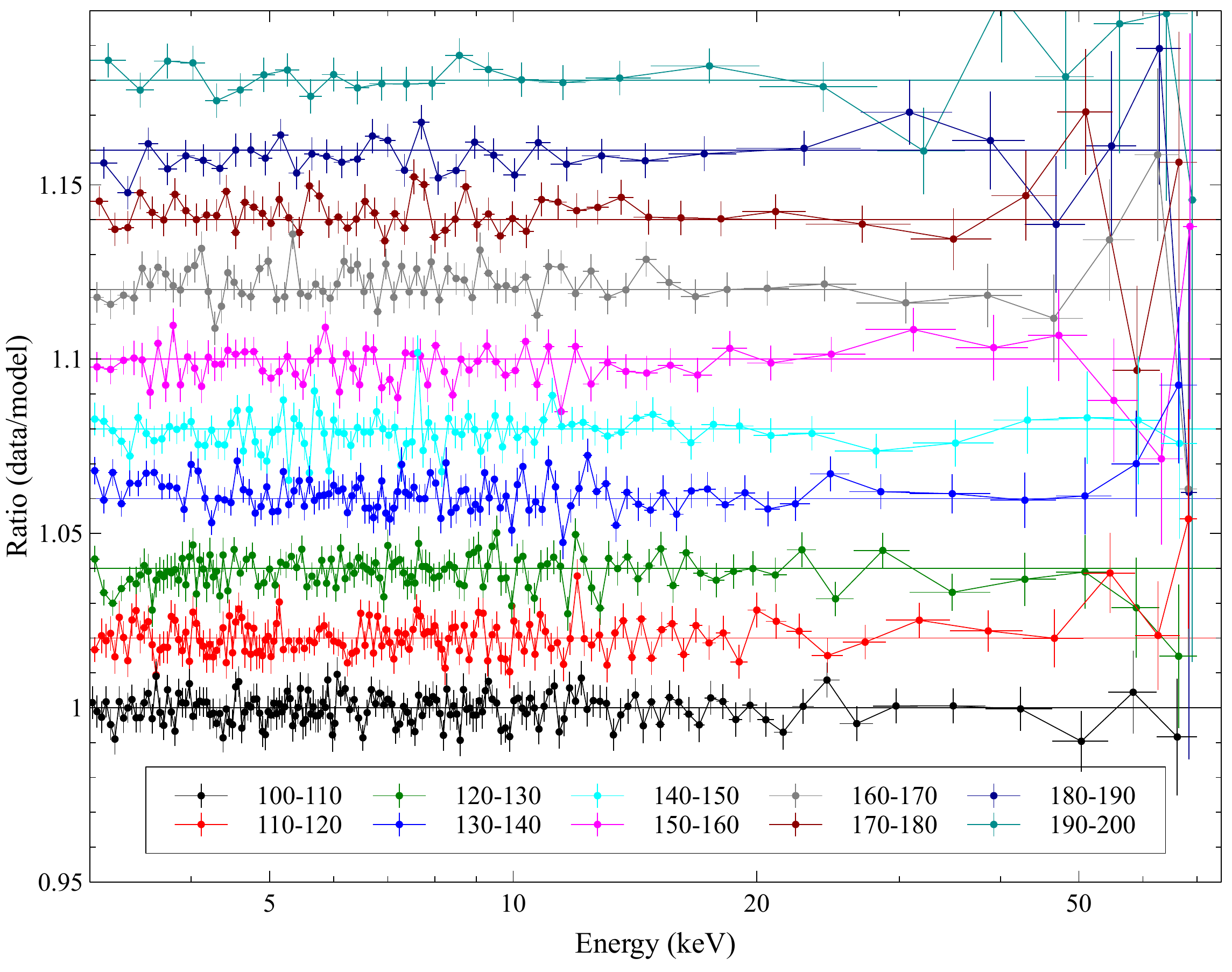}
\includegraphics[width=0.45\textwidth]{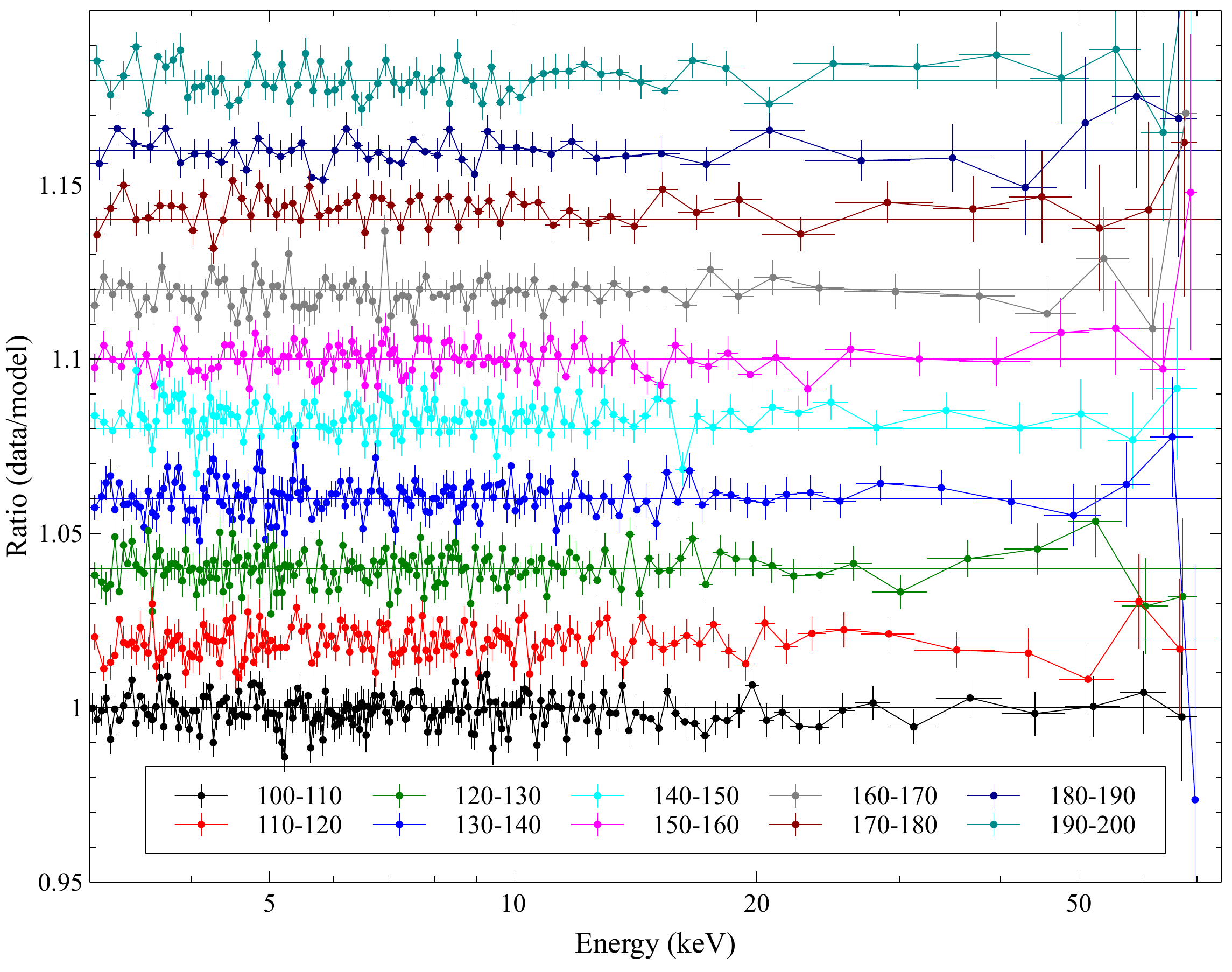}
\end{center}
\caption{Residual spectra for 10\as\ and 60\as\, artificially offset from one another for clarity. Separation between curves are 0.02 for 10\as\ and 0.03 for 60\as. These can be compared to Figure \ref{ratio1am} and Figure \ref{ratio10as}.}
\label{correctionspectra}
\end{figure}

We fitted each 10\as\ from 0\as out to 420\as. Because each off-axis bin has different fidelity, the process required supervision by adjusting the binning and ranges to ensure good results. The general guidelines listed here were followed whenever possible:
\begin{itemize}
    \item For an off-axis angle (oaa) between 1--4\am, the entire energy range from 2.2--78~keV was used.
    \item For off-axis angles $\theta <$1\am\ and 4\am$<\theta$, the upper energy range for the fit was gradually reduced from 80~keV to 40, 20, and 10~keV, depending on how good the signal was at those energies. 
    \item For off-axis angles that required an upper energy range decrease, we first fitted the nearest 1\am\ data set (see Figure \ref{correctionspectra}) and applied the resulting fit to the 10\as\ dataset. If by inspection the fit was tolerable to the data points above the upper energy range cut-off for the 10\as\ dataset,  the 1\am\ fit was used for the data above without a refit, while the data below were fitted as normal.
\end{itemize}

Figure \ref{correctionspectra} shows the spectra after correction with respect to the canonical reference Crab spectrum for 1\am\ bins (top panel) and for 10\as\ bins (bottom panel) for off-angle bins 100--200 \as. The residuals are on the order of $\pm$ 2\% below 50~kev, while above 50~keV there are larger deviation. The Crab is a soft X-ray source and, despite its brightness, the signal at high energies is limited, in particular, for higher off-axis angles and 5-10\% residuals should be tolerated.

\section{Results}\label{result}
\subsection{Crab Results}
\begin{figure}
\begin{center}
\includegraphics[width=0.95\textwidth]{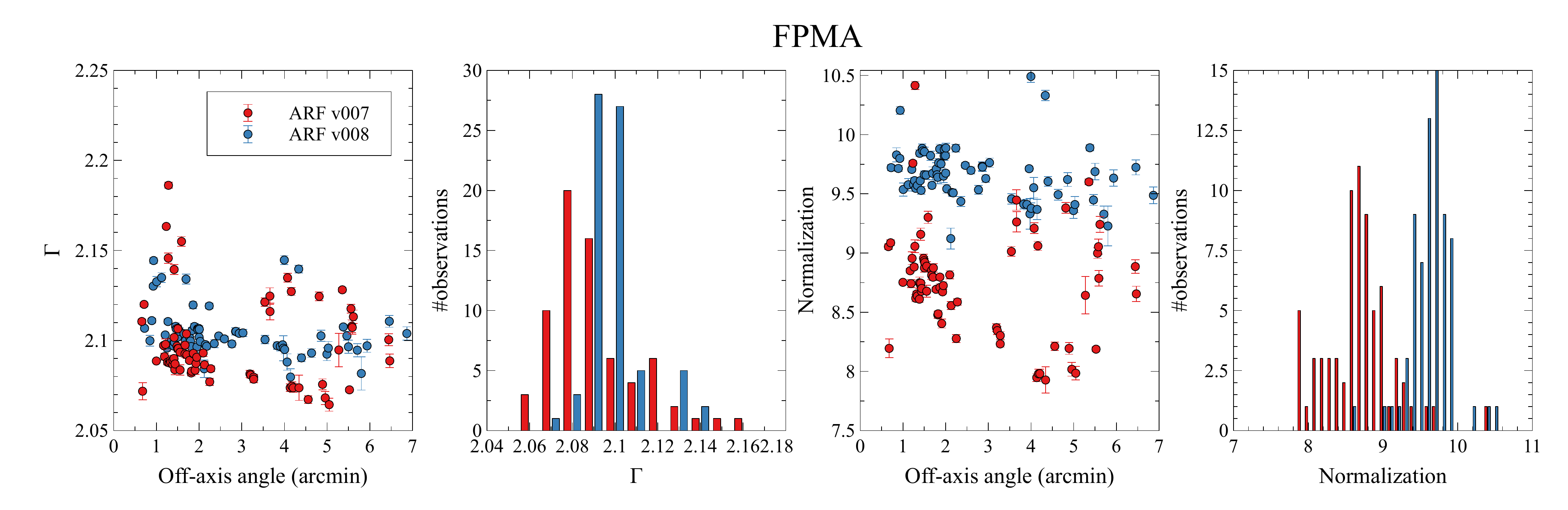}
\includegraphics[width=0.95\textwidth]{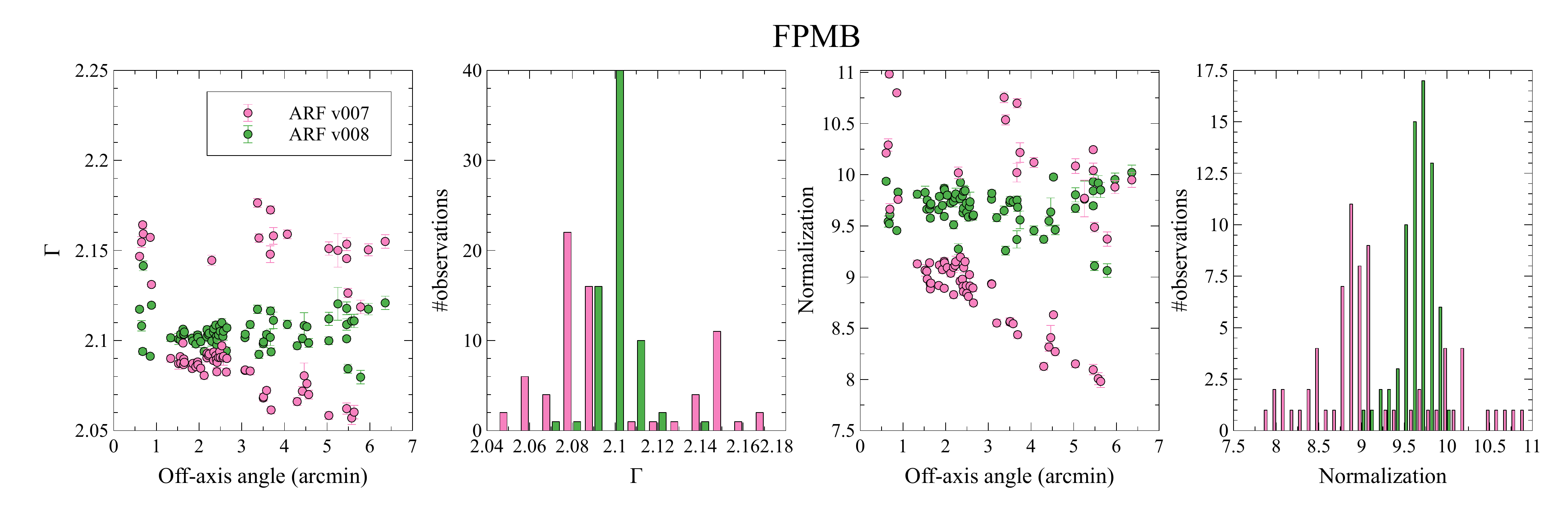}
\end{center}
\caption{Fitting results from all focused Crab observations from Table \ref{focusobslog} were reduced with ARF v007 and v008. For the powerlaw slope the spread in values is significantly improved above 3\am\ off-axis angle. Scatter still persists below 1\am\ due to the pulsar falling into the detector gaps, resulting in a change of the spectral parameters. For the normalization we observe a clear trend as a function of off-axis angle for the old responses which has been corrected in the new responses. The resulting flux increase in reprocessed observations is therefore a function of the off-axis angle of the observation. The statistical average values are recorded in Table \ref{AcompareB}.} 
\label{crabhistogram}
\end{figure}

To quantify and validate the new responses (\detabs\ v004, RMF v3.1, and ARF v008), we applied them to all of the focused Crab observations and compared the data to the old responses (\detabs\ v003, RMF 2010v002, ARF v007). The data was reduced using NUSTARDAS with the same conditions and parameters as discussed in Section \ref{reduction}. We fit between 3--70~keV, using the same absorbed powerlaw as before but without \texttt{nuabs}, which is now applied to the responses in \texttt{NUSTARDAS}. The results are shown in Figure \ref{crabhistogram}, and they demonstrate that we have succeeded in removing a strong dependency on off-axis angle in both slope and normalization that was present in the old responses. There still remains a minor residual slope at small and large off-axis angles, which are related to the source falling into the detector gaps at small off-axis angles, while for large off-axis angles there are complexities in the single-reflection component, which merges with the double-reflection PSF, along with the source being partially off the detector. Overall, though, for both modules the spread in slope and normalization has been significantly reduced in the new responses.

\begin{figure}
\begin{center}
\includegraphics[width=0.95\textwidth]{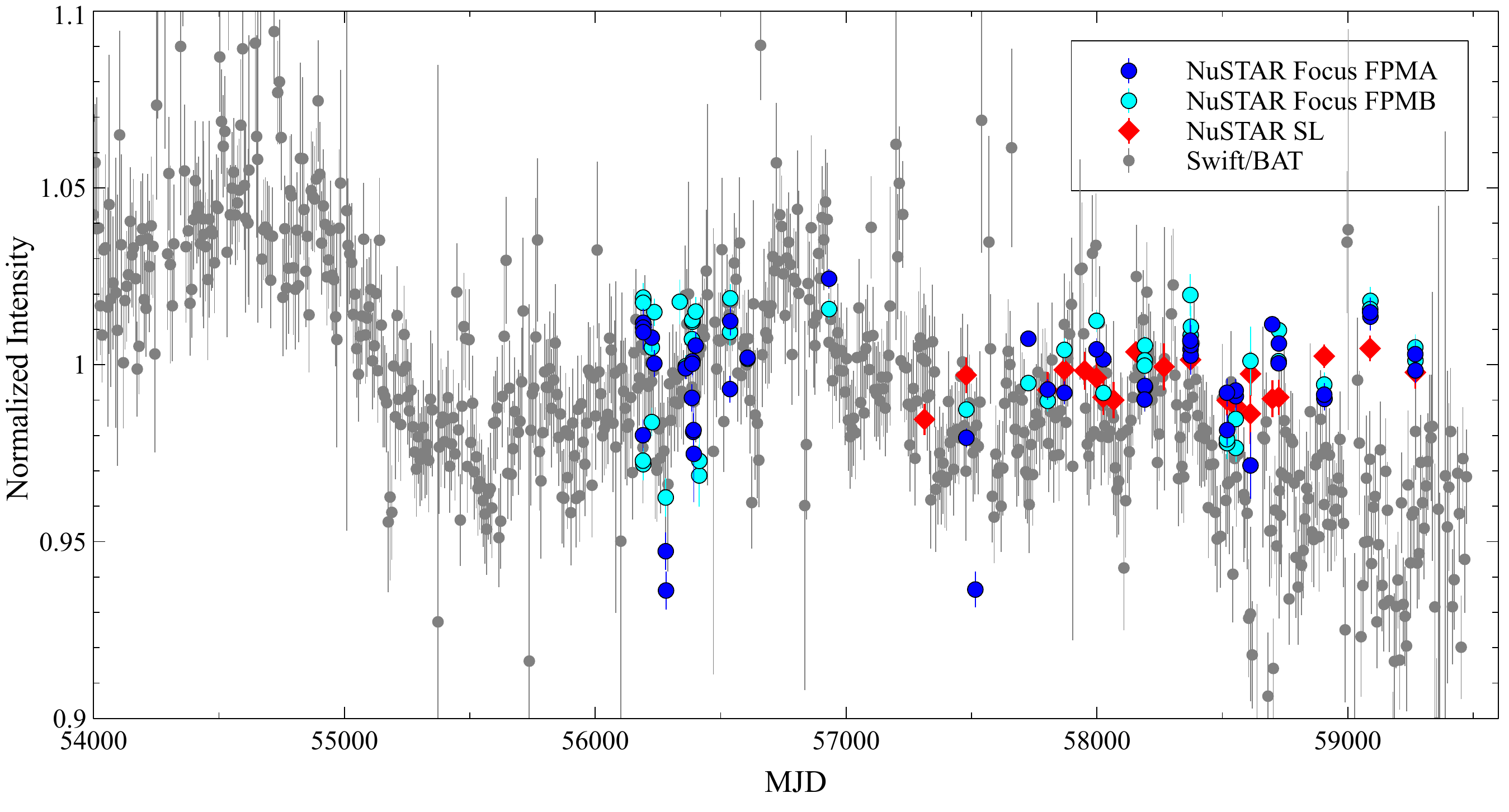}
\end{center}
\caption{Swift BAT 15-50~keV flux Crab lightcurve with \nustar\ SL and focused fluxes normalized to the 3--10~keV flux of the canonical Crab model used in this paper. The normalization between \nustar\ and Swift is relative and adjusted to be =1 at the start of the \nustar\ Crab observations. For the \nustar\ focused data we only show observations between 1-4~\am off-axis.}
\label{BATlc}
\end{figure}

The question of how much of the residual structure is due to calibration and how much due to the intrinsic variations of the Crab, still remains. To address this, we show in Figure \ref{BATlc} a lightcurve in the 15--50~keV band from Swift/BAT with the fluxes of the \nustar\ focused and SL observations used here. The \nustar\ data has been normalized to the canonical flux between 3--10~keV, and the Swift/BAT data adjusted to be $\sim$1 at the beginning of the \nustar\ Crab observations. We also calculated the \nustar\ intensity in the 15--50~keV band directly from the countrates by subtracting the background and dividing with the response in each channel before summation, and it yielded exactly the same relative intensities as the fluxes derived from the model fit. 

The long term Swift/BAT lightcurve shows a slow variation in the Crab that can be on the order of 1--2\% per year, however, the \nustar\ data does not appear to be observing this trend towards late times when Swift/BAT is detecting a decrease. It is important to understand that although we have calibrated to a constant flux and slope, the re-calibrated responses do not have a time dependency. The only two time dependent components in \texttt{NUSTARDAS} is the gain, which is linear, and the MLI correction, which is step-wise\cite{Madsen2020}. Since neither of these components have been calibrated with respect to the Crab, there is no reason to suspect that they would counteract the observed Crab variations, and it remains unclear why the two observatories are not tracking the same variation. What we can say, is that the re-calibrated \nustar\ data varies \textit{less} than what is observed with Swift/BAT. 

\begin{figure}
\begin{center}
\includegraphics[width=0.95\textwidth]{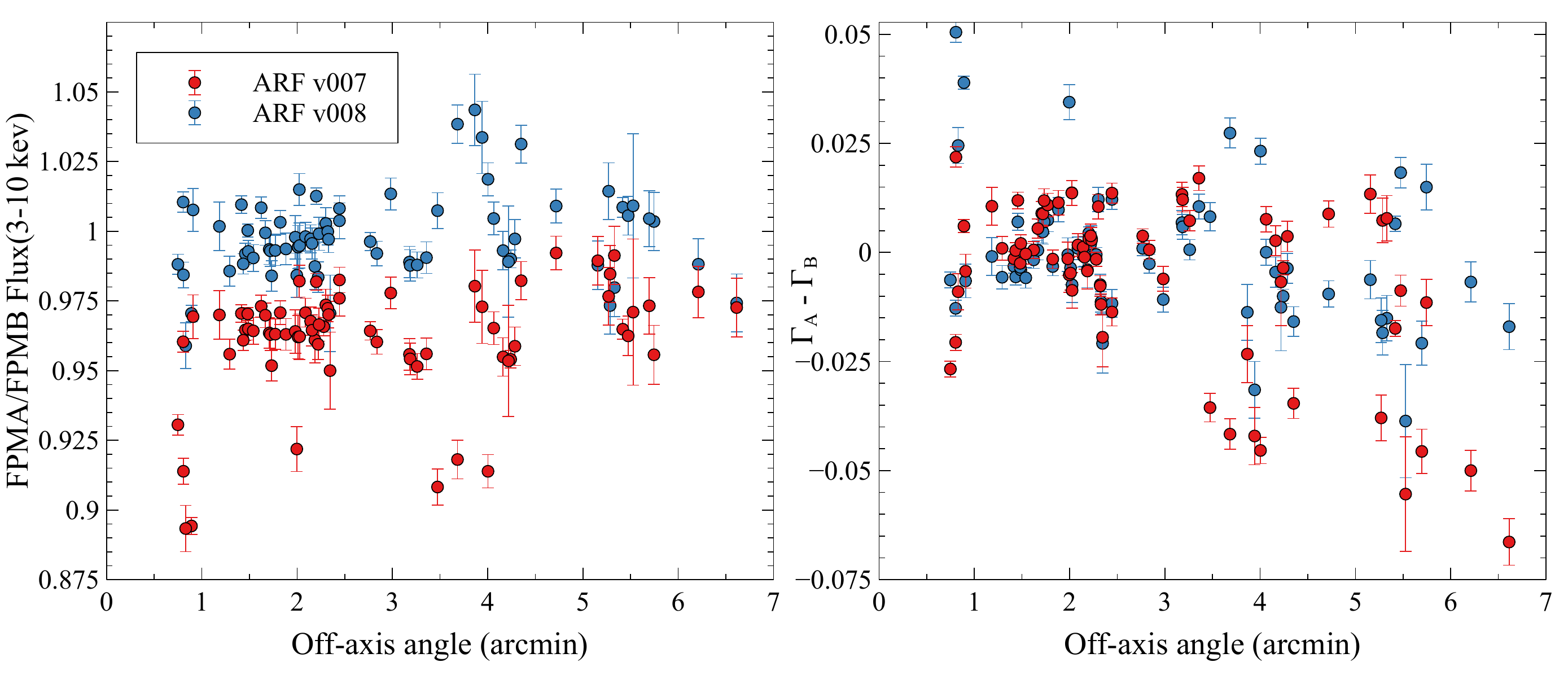}
\end{center}
\caption{Lef: Flux ratio between FPMA and FPMB for 3--10~keV, demonstrating that overall the two modules are in better agreement with the new responses. The scatter is unchanged, which indicates it is driven by secondary geometrical corrections like the PSF corrections and detector gaps. Right: Delta of the powerlaw index between FPMA nd FPMB. Improvements can be seen for high off-axis angles.}
\label{compareaAB}
\end{figure}

Since the remaining variations in teh \nustar\ data do not appear to be driven by the Crab, we can consider the distribution of fluxes and slopes across epochs for the focused Crab observations to be a good standard of our systematic error of repeated measurements. Table \ref{AcompareB} summarizes the mean and standard deviations of the $\Gamma$, Normalization, and the 3--10~keV flux for the focused observations with both the old and new responses. To avoid any bias from observations from low and high off-axis angles, we only evaluated the observations between 1\am--4\am, and in conclusion the systematic errors on the power-law slope is $\pm 0.01$. This should be taken in conjunction with Figure \ref{modelplot}, which bounds the possible errors on the model components. It is a little harder to place constraints on the flux, but Figure \ref{BATlc} and \ref{compareaAB} should serve as a guide and $\pm 3\%$ taken as a conservative estimate.

Finally, we investigated the FPMA/FPMB ratio and show in Figure \ref{compareaAB} that with the new responses as a whole the distribution of ratios has been shifted and centered about 1. The spread, however, largely remains the same, e.i the difference between FPMA and FPMB with repeated observations has not significantly changed. We interpret this to mean that the remaining errors are driven by secondary geometrical corrections, such as detector gaps, PSF correction, dead pixels, and variations of \detabs\ and detector thresholds within the detector.

\begin{table}
\centering
\caption{Mean and standard deviation of all focused Crab observations between 1\am--4\am.}\label{AcompareB}
\begin{tabular}{c|ccc}
\hline
Module  & $\Gamma$ & Normalization & Flux (3--10~keV) \\
& & (keV$^{-1}$ cm$^{-2}$ s$^{-1}$) & ($10^{-8}$~ergs~cm$^{-2}$~s$^{-1}$) \\
\hline
\hline
\multicolumn{4}{c}{ARF v007}\\
\hline
Crab$^\dagger$ & 2.1 & 8.5 & 1.38 \\
FPMA & 2.097 $\pm$     0.023 & 8.74 $\pm$      0.44 & 1.42 $\pm$     0.033 \\
FPMB & 2.103 $\pm$     0.033 & 9.20 $\pm$      0.71 & 1.49 $\pm$     0.042 \\
\hline
\multicolumn{4}{c}{ARF v008}\\
\hline
Crab$^\dagger$ & 2.013 & 9.69 & 1.57 \\
FPMA & 2.104 $\pm$ 0.013 & 9.66 $\pm$  0.23 & 1.56 $\pm$ 0.026 \\
FPMB & 2.105 $\pm$ 0.009 & 9.68 $\pm$  0.19 & 1.56 $\pm$ 0.028 \\
\hline
\multicolumn{4}{l}{$\dagger$ Canonical Crab values used for the calibration.}
\end{tabular}
\end{table}

\subsection{Case Studies}
\label{sec:case_studies}
To further help clarify the changes, we present here a couple of case studies where we have applied the new \texttt{CALDB} to some published and public data spanning a range of brightness and source types.

\subsubsection{TDE candidate SDSSJ143359.16+400636.0}
The observation took place on 2020 February 13 (Obdsid 90601606002) and published in Brightman et al. (2021) \cite{Brightman2021}. The spectrum is quite soft ($\Gamma\sim$3) and relatively faint (count rate $\sim$0.1 cts/s). The original analysis was done jointly with 2 Chandra data sets and we included that data in the reanalysis too. The model used was a simple \texttt{const $\times$ tbabs $\times$ powerlaw}. Observed changes to these parameters were:
\begin{itemize}
\item \nh: ($10^{20} \mathrm{cm}^{-2}): 9.4 \pm 5.3 \rightarrow 8.1 \pm 5.3$
\item $\Gamma: 2.90 \pm 0.11 \rightarrow 2.86 \pm 0.11$
\item Norm: $1.16 \pm 0.20 \times 10^{-3} \rightarrow 1.15 \pm 0.20 \times 10^{-3}$
\item Flux (erg cm$^{-2}$ s$^{-1}$, 3--15~keV): $5.83 \pm 0.28 \times 10^{-13} \rightarrow 6.16 \pm 0.33 \times 10^{-12}$
\end{itemize}
Most of the parameter changes are on the 1–2\% level, and much less than the uncertainties, with the exception of \nh, but as discussed the sensitivity to \nh\ in the \nustar\ band appoximately $1 \times 10^{21} \mathrm{cm}^{-2}$. The change in flux is +5\% as anticipated. This has no significant implications on the interpretation of the data.

\subsubsection{Ultra-compact X-ray binary 4U 1543-624}
Joint \nustar\ and NICER observations of the ultracompact X-ray binary (UCXB) 4U 1543-624 (Obsid 30601006002) were obtained in 2020 April, and the data published in Ludlam et al. (2021)\cite{Ludlam2021}. The comparison was made for the \nustar\ data only, and just for the continuum modeling using the model: \texttt{constant $\times$ TBfeo $\times$ (bbody + cutoffpl)}. The refitted parameter values agree within the 90\% confidence level, and the model flux (3--40~keV):
\begin{itemize}
    \item FPMA: $3.6098 \times 10^{-10} \rightarrow 3.8952 \times 10^{-10}$ erg cm$^{-2}$ s$^{-1}$ ($\sim$8\% increase)
    \item FPMB: $3.6511 \times 10^{-10} \rightarrow 3.8479 \times 10^{-10}$ erg cm$^{-2}$ s$^{-1}$ ($\sim$5\% increase)
\end{itemize}
Apart from the flux change there are no implications on the interpretation of the data.


\subsubsection{Black Hole X-ray Binary 4U 1543-475}
\begin{figure}
\begin{center}
\includegraphics[width=0.80\textwidth]{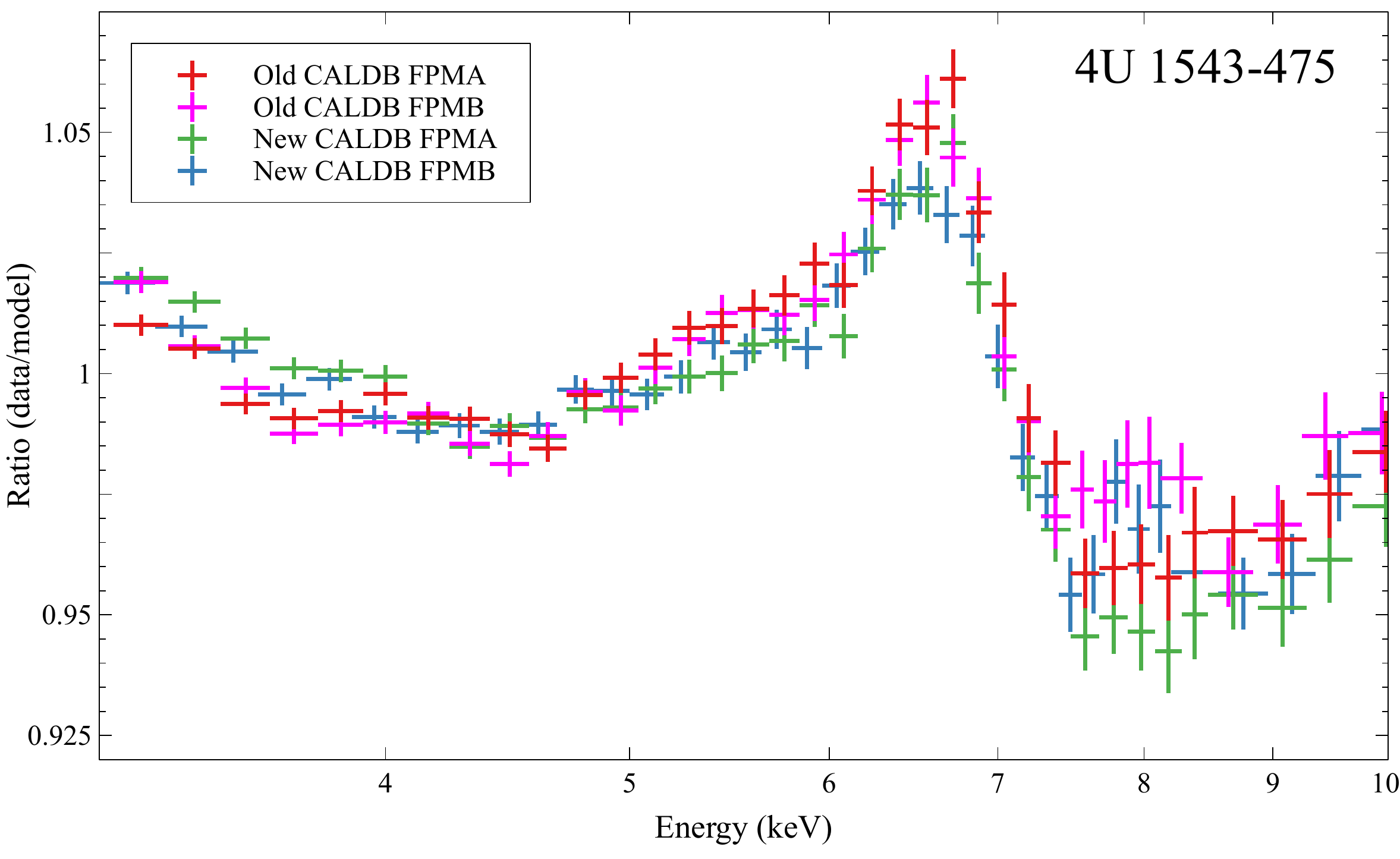}
\end{center}
\caption{Ratio plot of data extracted with v007 and v008 ARFs demonstrating the slope shift in the continuum.}
\label{BHB4u}
\end{figure}

 To evaluate the impact on more complex spectra, we investigated the bright black hole binary 4U 1543-475. This observation of 4U 1543-475 (obsid 90702326008) was a DDT trigger performed on August 29, 2021. The source count rate in the 3-79 keV band was roughly 400 cts/s, resulting in about 3 million total counts per FPM. We show the ratios of a simple \texttt{disk+powerlaw} fit to the before and after (with only a cross-normalization constants free between the model) in Figure \ref{BHB4u}, and this reveals the slope change introduced by the new responses. To compare how this slope change affects the source parameters, we used the model: \texttt{const $\times$ TBabs(simplcut $\times$ diskbb + relxillNS + xillverCp)}. 
 
 There is a flux increases in FPMA by $\sim$9\%, FPMB by $\sim$6\%, and this is captured almost entirely by an increase in disk normalization of a similar order ($\sim$10\%). The shift in spectral slope is more difficult to identify in only one parameter, and it is likely captured by small readjustments of the different model components in the soft band (e.g., \texttt{relxillNS} changes its normalization). 
 
\subsubsection{Cygnus X-1}
The observation was performed on August 2018 (obsID 80502335002), and the model used here for the comparison is: \texttt{constant $\times$ TBabs $\times$ (diskbb + cutoffpl + relxill)}, and the observed differences for the model are:
\begin{itemize}
    \item $T_{in}$ (temperature of \texttt{diskbb} model) : $0.41_{-0.03}^{+0.04} \rightarrow 0.53_{-0.05}^{+0.07}$ 
    \item \texttt{diskbb} norm : $11228_{-6107}^{+12993} \rightarrow 2185_{-1100}^{+2157}$
    \item $\Gamma: 1.74_{-0.03}^{+0.01} \rightarrow 1.71 \pm 0.02$
    \item Iron abundance: $4.6_{-0.2}^{+0.9} \rightarrow 5.7 \pm 1.0$
    \item Inclination (deg): $40_{-3}^{+1} \rightarrow 34_{-1}^{+3}$
\end{itemize}
We note that the slope and flux difference in the data set, which is induced by the differences in the responses, is once again mostly compensated for in \texttt{diskbb}.

\section{Conclusion}
We have in this paper presented a new calibration for \nustar\ released in the CALDB update 20211020. This release includes changes to the vignetting function for both telescopes, the detector absorption parameters, and the RMF of all eight detectors. The updated responses were calibration against the Crab spectral parameters, $\Gamma \equiv 2.103$ and N$ \equiv 9.69$, obtained from \nustar\ stray-light observations, and are a change from the previous calibration performed against $\Gamma \equiv 2.1$ and N$ \equiv 8.5$. This results in a model flux increase of 14\%, however, due to an off-axis angle dependency to the flux in the previous calibration, the increase observed is 5--15\%. The new responses significantly improve the response for high off-axis angles (\textgreater~4\am) and better agreement between the fluxes and slopes measured by FPMA and FPMB. The systematic errors for power-law slope and flux for repeated observations between 1\am--4\am\ is $\sigma_\Gamma = 0.01$ and for the flux in 3--10~keV $\sigma_\mathrm{flux} = 0.03 \times 10^{-8}$ erg cm$^{-2}$ s$^{-1}$.

We summarize here a set of general changes that may be seen when applying the new CALDB to old data:
\begin{itemize}
\item[1] A flux increase of 5-15\% depending on off-axis angle.
\item[2] The new RMFs  have been calibrated down to 2.2 keV to ensure that the data at 3~keV is resistant to redistribution errors from lower energy, \textbf{but data should still only be fitted down to 3.0~keV.}
\item[3] Significant effective area changes can move power-law slopes around by 0.04, again depends on off-axis angle.
\item[4] More accurate high-energy and high off-axis angle corrections.
\item[5] Better FPMA v. FPMB agreement in flux.
\end{itemize}

The re-analysis of several observations with the new calibration most strongly reflects the flux change, and though the new responses caused a minor reshuffling of fitted parameters, none resulted in a new interpretation of the data. 

\acknowledgments 
This work was supported under NASA Contract No. NNG08FD60C, and made use of data from the NuSTAR mission, a project led by the California Institute of Technology, managed by the Jet Propulsion Laboratory, and funded by the National Aeronautics and Space Administration. This research has made use of the NuSTAR Data Analysis Software (NuSTARDAS) jointly developed by the ASI Space Science Data Center (SSDC, Italy) and the California Institute of Technology (USA). This material is based upon work supported by NASA under award number 80GSFC21M0002. We thank Murray Brightman, Renee Ludlam, Guglielmo Mastroserio, and Riley Connors for performing the case study re-analyses in \S \ref{sec:case_studies}.
 
\bibliography{bib} 

\begin{thebibliography}{10}

\bibitem{Harrison2013}
{Harrison}, F.~A., {Craig}, W.~W., {Christensen}, F.~E., {Hailey}, C.~J.,
  {Zhang}, W.~W., {Boggs}, S.~E., {Stern}, D., {Cook}, W.~R., {Forster}, K.,
  {Giommi}, P., {Grefenstette}, B.~W., {Kim}, Y., {Kitaguchi}, T., {Koglin},
  J.~E., {Madsen}, K.~K., {Mao}, P.~H., {Miyasaka}, H., {Mori}, K., {Perri},
  M., {Pivovaroff}, M.~J., {Puccetti}, S., {Rana}, V.~R., {Westergaard}, N.~J.,
  {Willis}, J., {Zoglauer}, A., {An}, H., {Bachetti}, M., {Barri{\`e}re},
  N.~M., {Bellm}, E.~C., {Bhalerao}, V., {Brejnholt}, N.~F., {Fuerst}, F.,
  {Liebe}, C.~C., {Markwardt}, C.~B., {Nynka}, M., {Vogel}, J.~K., {Walton},
  D.~J., {Wik}, D.~R., {Alexander}, D.~M., {Cominsky}, L.~R., {Hornschemeier},
  A.~E., {Hornstrup}, A., {Kaspi}, V.~M., {Madejski}, G.~M., {Matt}, G.,
  {Molendi}, S., {Smith}, D.~M., {Tomsick}, J.~A., {Ajello}, M., {Ballantyne},
  D.~R., {Balokovi{\'c}}, M., {Barret}, D., {Bauer}, F.~E., {Blandford}, R.~D.,
  {Brandt}, W.~N., {Brenneman}, L.~W., {Chiang}, J., {Chakrabarty}, D.,
  {Chenevez}, J., {Comastri}, A., {Dufour}, F., {Elvis}, M., {Fabian}, A.~C.,
  {Farrah}, D., {Fryer}, C.~L., {Gotthelf}, E.~V., {Grindlay}, J.~E.,
  {Helfand}, D.~J., {Krivonos}, R., {Meier}, D.~L., {Miller}, J.~M.,
  {Natalucci}, L., {Ogle}, P., {Ofek}, E.~O., {Ptak}, A., {Reynolds}, S.~P.,
  {Rigby}, J.~R., {Tagliaferri}, G., {Thorsett}, S.~E., {Treister}, E., and
  {Urry}, C.~M., ``{The Nuclear Spectroscopic Telescope Array (NuSTAR)
  High-energy X-Ray Mission},'' {\em \apj}~{\bf 770},  103 (June 2013).

\bibitem{Petre1985}
{Petre}, R. and {Serlemitos}, P., ``{Conical Imaging Mirrors for High-Speed
  X-ray Telescopes},'' {\em Applied Optics}~{\bf 24},  1833--1837 (1985).

\bibitem{Madsen2009}
{Madsen}, K.~K., {Harrison}, F.~A., {Mao}, P.~H., {Christensen}, F.~E.,
  {Jensen}, C.~P., {Brejnholt}, N., {Koglin}, J., and {Pivovaroff}, M.~J.,
  ``{Optimizations of Pt/SiC and W/Si multilayers for the Nuclear Spectroscopic
  Telescope Array},'' in [{\em Proc. SPIE}{\nolinebreak\hspace{0.1em}]},   {\bf
  7437} (Aug. 2009).

\bibitem{Madsen2015a}
{Madsen}, K.~K., {Harrison}, F.~A., {Markwardt}, C.~B., {An}, H.,
  {Grefenstette}, B.~W., {Bachetti}, M., {Miyasaka}, H., {Kitaguchi}, T.,
  {Bhalerao}, V., {Boggs}, S., {Christensen}, F.~E., {Craig}, W.~W., {Forster},
  K., {Fuerst}, F., {Hailey}, C.~J., {Perri}, M., {Puccetti}, S., {Rana}, V.,
  {Stern}, D., {Walton}, D.~J., {J{\o}rgen Westergaard}, N., and {Zhang},
  W.~W., ``{Calibration of the NuSTAR High-energy Focusing X-ray Telescope.},''
  {\em \apjs}~{\bf 220},  8 (Sept. 2015).

\bibitem{Toor1974}
{Toor}, A. and {Seward}, F.~D., ``{The Crab Nebula as a calibration source for
  X-ray astronomy},'' {\em AJ}~{\bf 79},  995--999 (oct 1974).

\bibitem{Kirsch2005}
{Kirsch}, M.~G., {Briel}, U.~G., {Burrows}, D., {Campana}, S., {Cusumano}, G.,
  {Ebisawa}, K., {Freyberg}, M.~J., {Guainazzi}, M., {Haberl}, F., {Jahoda},
  K., {Kaastra}, J., {Kretschmar}, P., {Larsson}, S., {Lubi{\'n}ski}, P.,
  {Mori}, K., {Plucinsky}, P., {Pollock}, A.~M., {Rothschild}, R., {Sembay},
  S., {Wilms}, J., and {Yamamoto}, M., ``{Crab: the standard x-ray candle with
  all (modern) x-ray satellites},'' in [{\em Proc.
  SPIE}{\nolinebreak\hspace{0.1em}]},  {Siegmund}, O.~H.~W., ed.,  {\bf 5898},
  22--33 (Aug. 2005).

\bibitem{Madsen2017b}
{Madsen}, K.~K., {Forster}, K., {Grefenstette}, B.~W., {Harrison}, F.~A., and
  {Stern}, D., ``{Measurement of the Absolute Crab Flux with NuSTAR},'' {\em
  \apj}~{\bf 841},  56 (May 2017).

\bibitem{Madsen2020}
{Madsen}, K.~K., {Grefenstette}, B.~W., {Pike}, S., {Miyasaka}, H.,
  {Brightman}, M., {Forster}, K., and {Harrison}, F.~A., ``{NuSTAR low energy
  effective area correction due to thermal blanket tear},'' {\em arXiv
  e-prints} ,  arXiv:2005.00569 (May 2020).

\bibitem{Jourdain2009}
{Jourdain}, E. and {Roques}, J.~P., ``{The High-Energy Emission of the Crab
  Nebula from 20 keV TO 6 MeV with Integral SPI},'' {\em \apj}~{\bf 704},
  17--24 (Oct. 2009).

\bibitem{Weisskopf2010}
{Weisskopf}, M.~C., {Guainazzi}, M., {Jahoda}, K., {Shaposhnikov}, N.,
  {O'Dell}, S.~L., {Zavlin}, V.~E., {Wilson-Hodge}, C., and {Elsner}, R.~F.,
  ``{On Calibrations Using the Crab Nebula and Models of the Nebular X-Ray
  Emission},'' {\em \apj}~{\bf 713},  912--919 (Apr. 2010).

\bibitem{Wilson2011}
{Wilson-Hodge}, C.~A., {Cherry}, M.~L., {Case}, G.~L., {Baumgartner}, W.~H.,
  {Beklen}, E., {Narayana Bhat}, P., {Briggs}, M.~S., {Camero-Arranz}, A.,
  {Chaplin}, V., {Connaughton}, V., {Finger}, M.~H., {Gehrels}, N., {Greiner},
  J., {Jahoda}, K., {Jenke}, P., {Kippen}, R.~M., {Kouveliotou}, C., {Krimm},
  H.~A., {Kuulkers}, E., {Lund}, N., {Meegan}, C.~A., {Natalucci}, L.,
  {Paciesas}, W.~S., {Preece}, R., {Rodi}, J.~C., {Shaposhnikov}, N.,
  {Skinner}, G.~K., {Swartz}, D., {von Kienlin}, A., {Diehl}, R., and {Zhang},
  X.-L., ``{When a Standard Candle Flickers},'' {\em \apjl}~{\bf 727},  L40
  (Feb. 2011).

\bibitem{Shaposhnikov2012}
{Shaposhnikov}, N., {Jahoda}, K., {Markwardt}, C., {Swank}, J., and
  {Strohmayer}, T., ``{Advances in the RXTE Proportional Counter Array
  Calibration: Nearing the Statistical Limit},'' {\em \apj}~{\bf 757},  159
  (Oct. 2012).

\bibitem{Madsen2017a}
Madsen, K.~K., Christensen, F.~E., Craig, W.~W., Forster, K.~W., Grefenstette,
  B.~W., Harrison, F.~A., Miyasaka, H., and Rana, V., ``Observational artifacts
  of nuclear spectroscopic telescope array: ghost rays and stray light,'' {\em
  Journal of Astronomical Telescopes, Instruments, and Systems}~{\bf 3},  3 --
  3 -- 13 (2017).

\bibitem{Wilms2000}
{Wilms}, J., {Allen}, A., and {McCray}, R., ``{On the Absorption of X-Rays in
  the Interstellar Medium},'' {\em \apj}~{\bf 542},  914--924 (Oct. 2000).

\bibitem{Verner1996}
{Verner}, D.~A., {Ferland}, G.~J., {Korista}, K.~T., and {Yakovlev}, D.~G.,
  ``{Atomic Data for Astrophysics. II. New Analytic FITS for Photoionization
  Cross Sections of Atoms and Ions},'' {\em \apj}~{\bf 465},  487 (July 1996).

\bibitem{Grefenstette2021}
{Grefenstette}, B.~W., {Ludlam}, R.~M., {Thompson}, E.~T., {Garc{\'\i}a},
  J.~A., {Hare}, J., {Jaodand}, A.~D., {Krivonos}, R.~A., {Madsen}, K.~K.,
  {Mastroserio}, G., {Slaughter}, C.~M., {Tomsick}, J.~A., {Wik}, D., and
  {Zoglauer}, A., ``{StrayCats: A Catalog of NuSTAR Stray Light
  Observations},'' {\em \apj}~{\bf 909},  30 (Mar. 2021).

\bibitem{Cirrone2010}
{Cirrone}, G.~A.~P., {Cuttone}, G., {Di Rosa}, F., {Pandola}, L., {Romano}, F.,
  and {Zhang}, Q., ``{Validation of the Geant4 electromagnetic photon
  cross-sections for elements and compounds},'' {\em Nuclear Instruments and
  Methods in Physics Research A}~{\bf 618},  315--322 (June 2010).

\bibitem{grefenstette_2018_threshold}
Grefenstette, B.~W., Cook, W.~R., Harrison, F.~A., Kitaguchi, T., Madsen,
  K.~K., Miyasaka, H., and Pike, S.~N., ``{Pushing the limits of NuSTAR
  detectors},'' in [{\em High Energy, Optical, and Infrared Detectors for
  Astronomy VIII}{\nolinebreak\hspace{0.1em}]},  Holland, A.~D. and Beletic,
  J., eds.,  {\bf 10709},  705 -- 711, International Society for Optics and
  Photonics, SPIE (2018).

\bibitem{Madsen2015b}
{Madsen}, K.~K., {Reynolds}, S., {Harrison}, F., {An}, H., {Boggs}, S.,
  {Christensen}, F.~E., {Craig}, W.~W., {Fryer}, C.~L., {Grefenstette}, B.~W.,
  {Hailey}, C.~J., {Markwardt}, C., {Nynka}, M., {Stern}, D., {Zoglauer}, A.,
  and {Zhang}, W., ``{Broadband X-ray Imaging and Spectroscopy of the Crab
  Nebula and Pulsar with NuSTAR},'' {\em \apj}~{\bf 801},  66 (Mar. 2015).

\bibitem{Brightman2021}
{Brightman}, M., {Ward}, C., {Stern}, D., {Mooley}, K., {De}, K., {Gezari}, S.,
  {Van Velzen}, S., {Andreoni}, I., {Graham}, M., {Masci}, F.~J., {Riddle}, R.,
  and {Zolkower}, J., ``{A Luminous X-Ray Transient in SDSS
  J143359.16+400636.0: A Likely Tidal Disruption Event},'' {\em \apj}~{\bf
  909},  102 (Mar. 2021).

\bibitem{Ludlam2021}
Ludlam, R.~M., Jaodand, A.~D., Garc{\'{\i}}a, J.~A., Degenaar, N., Tomsick,
  J.~A., Cackett, E.~M., Fabian, A.~C., Gandhi, P., Buisson, D. J.~K., Shaw,
  A.~W., and Chakrabarty, D., ``Simultaneous {NICER} and {NuSTAR} observations
  of the ultracompact x-ray binary 4u 1543{\textendash}624,'' {\em The
  Astrophysical Journal}~{\bf 911},  123 (apr 2021).

\end{thebibliography}
\bibliographystyle{spiebib} 

\end{document}